\documentclass{IEEEtran}
\usepackage{cite}
\usepackage{amsmath,amssymb,amsfonts}
\usepackage{algorithmic}
\usepackage{graphicx}
\usepackage[caption=false]{subfig}
\usepackage{textcomp}
\usepackage{xcolor}
\usepackage{url}
\usepackage{bm} 
\def\BibTeX{{\rm B\kern-.05em{\sc i\kern-.025em b}\kern-.08emT\kern-.1667em\lower.7ex\hbox{E}\kern-.125emX}}

\begin{document}
	\title{
		Radiation and Scattering of EM Waves in Large Plasmas Around Objects in Hypersonic Flight
	}
	\author{A. Scarabosio, J. L. Araque Quijano, J. Tobon \IEEEmembership{Member, IEEE}, M. Righero, G. Giordanengo, D. D'Ambrosio, L. Walpot and G. Vecchi \IEEEmembership{Fellow, IEEE} 
	
	\thanks{A. Scarabosio, M. Righero and G. Giordanengo are with the LINKS Foundation, Via P. C. Boggio, 61 - Turin, Italy, (email: andrea.scarabosio@linksfoundation.com).}
	\thanks{J. Tobon, D. D'Ambrosio and G. Vecchi are with the Politecnico di Torino, Corso Duca degli Abruzzi, 24 - Turin, Italy.}
	\thanks{J. L. Araque Quijano is with the Universidad Nacional de Colombia, Bogota, Colombia.}
	\thanks{L. Walpot is with European Space Agency - ESA, Paris, France.}}
	
	\renewcommand{\Re}{\operatorname{Re}}
	\renewcommand{\Im}{\operatorname{Im}}	
	\maketitle
	
	\begin{abstract}
		Hypersonic flight regime is conventionally defined for Mach$>5$; in these conditions, the flying object becomes enveloped in a plasma. This plasma is densest in thin surface layers, but in typical situations of interest it impacts electromagnetic wave propagation in an electrically large volume. We address this problem with a hybrid approach. 
		We employ Equivalence Theorem to separate the inhomogeneous plasma region from the surrounding free space via an equivalent (Huygens) surface, and the Eikonal approximation to Maxwell equations in the large inhomogeneous region for obtaining equivalent currents on the separating surface. Then, we obtain the scattered field via (exact) free space radiation of these surface equivalent currents.
		The method is extensively tested against reference results and then applied to a real-life re-entry vehicle with full 3D plasma computed via Computational Fluid Dynamic (CFD) simulations. We address both scattering (RCS) from the entire vehicle and radiation from the on-board antennas.
		From our results, significant radio link path losses can be associated with plasma spatial variations (gradients) and collisional losses, to an extent that matches well the usually perceived blackout in crossing layers in cutoff. Furthermore, we find good agreement with existing literature concerning significant alterations of the radar response (RCS) due to the plasma envelope.
	\end{abstract}
	
	
	\section{Introduction and Motivation}\label{sec:introduction}
	Hypersonic flight regime is conventionally defined for Mach$>5$, also called hypervelocity regime; it is the standard condition of vehicles re-entering Earth atmosphere from various orbits, or entering extra-terrestrial atmospheres (e.g. Martian). It is also the physical condition of (proposed) future transport \cite{Hypersonic_civil_Boing_nbcnews,Hypersonic_civil_StratoFly}, of meteors (asteroids entering Earth atmosphere) \cite{Bariselli_ARC2018,Bariselli_AA_2020} and de-orbiting space debris (that must be monitored for safety). A very recent addition is the proposal, and in some case fielding, of hypersonic weapon systems \cite{Ross_Spectrum_2020}. 
	
	In these physical conditions, the extreme shock wave due to hypervelocity generates intense heat and ionization of atmospheric gases. When this happens, the flying object is enveloped in the resulting plasma, often with a significant trailing wake.
	
	This plasma may generate significant effects on the radio communication between the vehicle and ground or satellite relay links, as well as on access to Global Navigation Satellite Systems (GNSS). As well known, when the link direction crosses plasma layers with plasma frequency near or above the link frequency, a significant path loss emerges, usually called ``blackout''; lower levels of attenuation are often called ``brownout''. 
	Likewise, the plasma may significantly impact the radar return, in the first place its Radar Cross Section (RCS). 
	For meteor-type objects, the plasma also constitutes the ionized trailing tail, that allows its radar observation \cite{Bariselli_AA_2020}.   
	
	In typical conditions where the plasma has significant effects on radiation and scattering of EM waves, most flying objects and the plasma cloud around it (including the wake) are electrically large, and the plasma is inhomogeneous throughout.
	
	
	Full-wave methods have been reported for this problem. Volume integral equations can be accelerated, but pose significant issues of stability, addressed in \cite{Gomez_AWPL_2015,Yucel_TAP_2018}.  
	A discontinuous Galerkin method has been proposed in \cite{Wang_TAP_2020}; 
	the numerical examples in these relevant works have (correctly) focused on situations that were mostly critical for the respective methods; as a result, simple and relatively thin plasma layers have been considered there. 
	Finite Difference Time Domain (FDTD) can also be employed, with time convolution because of the associated dispersion; application to realistic re-entry cases is reported in \cite{Takahashi_Areo_2016}, requiring supercomputing facilities (45M unknowns). 
	
	In order to emulate situations happening in reality, the plasma distribution must come from 
	computational fluid dynamic (CFD) simulations; 
	these take as input the flying object shape and materials, the local atmosphere, and the flight trajectory; most relevant are the velocity and angle of attack. The trajectory may be known with accuracy for re-entering space vehicles, estimated for meteor-like objects, or largely unpredictable (which is conspicuously the case for hypersonic weapons \cite{Ross_Spectrum_2020}). Similar considerations can be made for the shape and materials of the hypersonic objects, with the additional observation that CFD boundary conditions depend on coating materials (e.g. ablative or non ablative thermal shields).
	
	Even when the trajectory is known with accuracy, and so are the geometry and materials of the object, uncertainties remain in the CFD predictions, as CFD boundary conditions are known from experimental datasets, and subject to Ansatz and assumptions. Finally, the possible onset of turbulence and its characteristic lengths may generate fluctuations in the electron densities with impact on the EM response.
	
	Overall, the plasma and hence its EM effect have to be considered to some degree stochastic \cite{Sotnikov_TPS_2010,Sotnikov_PP_2014}. On the most-deterministic end, one would like to consider error bars in the results, via a sensitivity analysis. In the most uncertain case, the results ideally should come from a Monte Carlo process. 
	
	As a result, the case emerges for approximate and numerically nimble simulation approaches; they can - inter alia - allow for parametric studies, ideally up to Monte Carlo characterization of the EM quantities of interest in the prediction endeavour.
	
	In this scenario, numerical methods based on asymptotic approximation of plasma propagation are good candidates to compromise between accuracy and computational load, as will be discussed in this work. 
	
	In this sense, our approach is inspired by the foundational work in \cite{Ling_TAP_1991,Kim92}. Those ideas, with some modifications and additions, are applied in this work to full 3D plasmas arising from CFD simulation of real flying objects; an extensive validation is performed prior to addressing EM characterization of a real-life atmospheric re-entry vehicle in hypersonic flight regime.  This real-life study case is the European Space Agency (ESA) IXV re-entry vehicle (shown in Fig. \ref{fig:model:IXV}); more information about the IXV vehicle and the related ESA mission can be found in \cite{DiBenedetto_JAE_2014,Angelini_AA_2016} and references therein. 
	
	The paper is organized as follows. In Sec. \ref{sec:model}
	we describe the physical and numerical model;  Sec. \ref{sec:benchmark} describes the validation of the method, and finally an application to an actual re-entry vehicle is reported in Sec. \ref{sec:IXV_plasma}.      
	
	Preliminary accounts of the present work were presented in conference publications 
	\cite{Vecchi_APS2004_plasmi,Bandinelli_EMC_2011,Pandolfo_ESA_2012}.
	
	\section{Physical and numerical model}\label{sec:model}
	
	\subsection{Plasma model: CFD model}\label{sec:cfd}
	The hypersonic flow about a re-entry vehicle is computed by solving the full Navier-Stokes equations for multi-component gas mixtures. The physical model includes species mass balance equations, and vibrational energy balance equations to account for high-temperature effects. The gas mixture is assumed to be composed of 11 species, namely O$_2$, N$_2$, O, N, NO, O$_2^+$, N$_2^+$, O$^+$, N$^+$, NO$^+$ and e$^-$. Chemical reaction rates are taken from \cite{Park_JTHT_7(3)_1993,ParkJaffePartridge_JTHT_15(1)_2001}, and the transport model is based on the Champman-Enskog method and it adopts the collisional integrals suggested in \cite{LevinWright_JTHT_18(1)_2004,WrightLevin_JTHT_19(1)_2005,Wright_et_al_AIAAJ_43(12)_2005,WrightLevin_JTHT_19(1)_2005,Gupta_NASA_1990}. The governing equations are discretised according to a finite volumes method. Convective fluxes are evaluated using standard upwind flux-difference splitting techniques, and diffusive fluxes through a centered scheme. Second order accuracy in space and in time is achieved using an Essentially Non-Oscillatory scheme. Chemical source terms are treated implicitly to avoid instabilities due to fast chemical reactions.
	
	\begin{figure}[!ht]
		\centerline{\includegraphics[width=0.95\columnwidth]{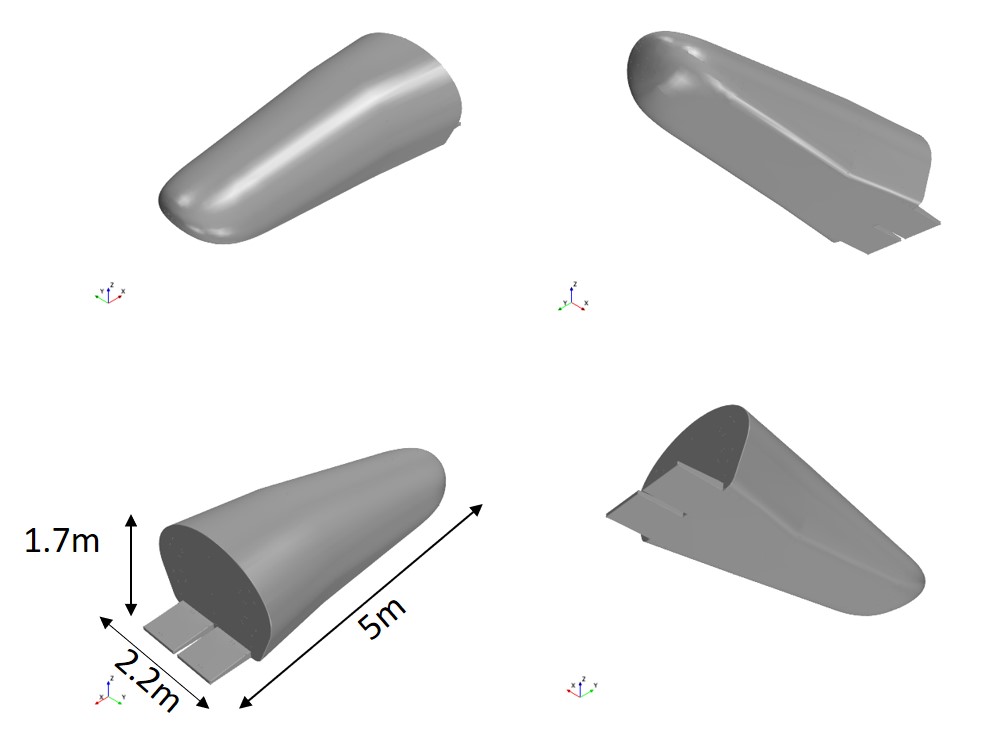}}
		\caption{CAD rendering of the ESA IXV re-entry vehicle.}
		\label{fig:model:IXV}
	\end{figure}
	
	\begin{figure}[!ht] 
		\centerline{\includegraphics[width=\columnwidth]{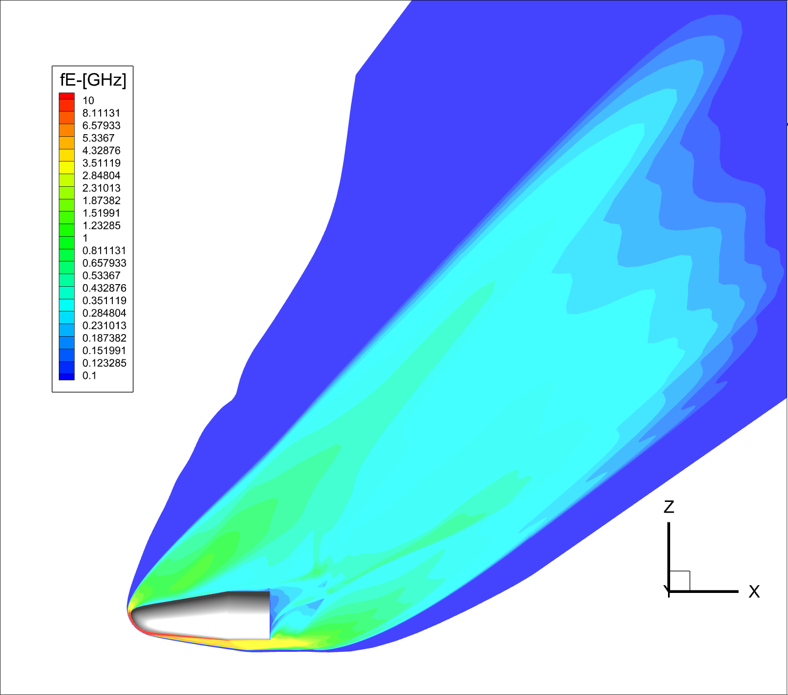}}      
		\caption{Plasma around the IXV vehicle during re-entry flight at Mach number 15, the figure depicts density as plasma frequency.}
		\label{fig:model:IXV_plasma}
	\end{figure}
	
	In the following, the plasma and gas thermodynamic quantities are calculated according to the model in \cite{DAmbrosio_THT_2007,DAmbrosio_AIAA_2009}. The model accounts for gas ionisation by shock waves due to the hypersonic flow. We stress the fact that these first principle calculations are already quite realistic and constitute the state-of-the-art for shock waves generated plasmas simulations. 
	
	
	A typical result, for an actual re-entry vehicle is shown in figure \ref{fig:model:IXV_plasma}. 
	

	\subsection{Plasma model: EM model}\label{sec:plasma_model}
	Throughout this work the medium is supposed to be stationary; a time-dependence  $ \exp(-i\omega t)$ in the EM fields is assumed and suppressed, with $\omega=2\pi f$ being the angular frequency at frequency $f$.
	
	We are interested in the frequency range where the plasma response is dominated by the electron motion; we do consider collisional effects in the plasma and neutral particles, but as the the wave phase velocity is much greater than plasma thermal velocity, temperature  and higher order collective kinetic effects can be left out. This is the so-called collisional unmagnetised ``cold plasma'' approximation in standard literature. In this regime the dielectric tensor is diagonal and the proper plasma constitutive relation is \cite{Stix_book}:
	\begin{equation}
		\epsilon_r(\bm{r})=1 -\frac{\omega_{pe}^2(\bm{r})}{\omega \left( \omega-i\nu_c(\bm{r})\right) } 
		\label{eq:plasma_dispersion}
	\end{equation}
	where $\omega_{pe}^2=n_ee^2/(\epsilon_0m_e)$ is the electron plasma frequency (squared), $n_e$ the electron density, $m_e$ the electron mass, $\epsilon_0$ the vacuum permittivity and $\nu_c$ is the neutral-plasma collision frequency. 
	Equation (\ref{eq:plasma_dispersion}) is also equivalent to well-known Appleton-Hartree equation (in the special case of no magnetic field), which has been extensively applied to ionospheric propagation \cite{Budden_1961}.
	For multi species partially ionised plasmas the collision frequency is related to temperature via \cite{Landau_StatPhys}:
	\begin{equation}
		\nu_c(\bm{r})=\Sigma_{\alpha}n_{\alpha,0}(\bm{r})\sigma_{\alpha,c}(T_e(\bm{r}))\sqrt{\frac{8k_B T_e(\bm{r})}{\pi m_e}}  
		\label{eq:collisional_freq}
	\end{equation}
	with $T_e$ being the electron temperature, $n_0$ the neutral density, $k_B$ the Boltzmann constant and $\sigma_{\alpha,c}$ is the neutral-electron scattering cross section for the neutral specie $\alpha$. The collisional frequency is a function of $T_e$ through both the electron mean velocity and the scattering cross section. Accurate fits of cross sections data, relevant for a hot air plasma, can be found in \cite{Gupta_NASA_1990} and are used here for the calculation of $\nu_c$. In a single ideal gas approximation a representative value of $\sigma_c$ for hot air is $1\cdot 10^{-19}$ m$^{-2}$ \cite{Lankford_1972,Gupta_NASA_1990}.     
	
	
	In our model, the spatial distribution of the hypersonic flow and of the related plasma parameters are known in numerical form from CFD simulations. Hence, data representation must be made consistent with the EM model. As detailed in Sec. \ref{sec:model:raypropagation}, 
	we need to compute both the relative electric permittivity and its gradient for a very high number of spatial locations; due to data origin, the gradient has to be computed numerically. If not done with care, this step may lead to inaccurate EM field calculation and/or very high computational cost. Moreover, the plasma profiles are typically computed in CFD simulation on an unstructured spatial grid. Most of the accurate and fast multivariate interpolation methods are available for regular gridded data set only, which makes the gradient calculation a critical data processing. We choose to process the CFD 3D plasma profiles in two steps. First, we interpolate $\epsilon_r(\bm{r})$ on a regular rectangular grid with typical resolution of $\approx\lambda/5$. Second, we fit the regular grid using a global cubic spline, that allows a rapid and smooth evaluation of the gradient, which is continuous by construction. This process is performed only once and the spine fitting is used to computed the r.h.s. of eq. \ref{eq:ray_equation2} during trajectory integration (see section \ref{sec:model:raypropagation}).

	\begin{figure}[!ht] 
		\centerline{\includegraphics[width=\columnwidth]{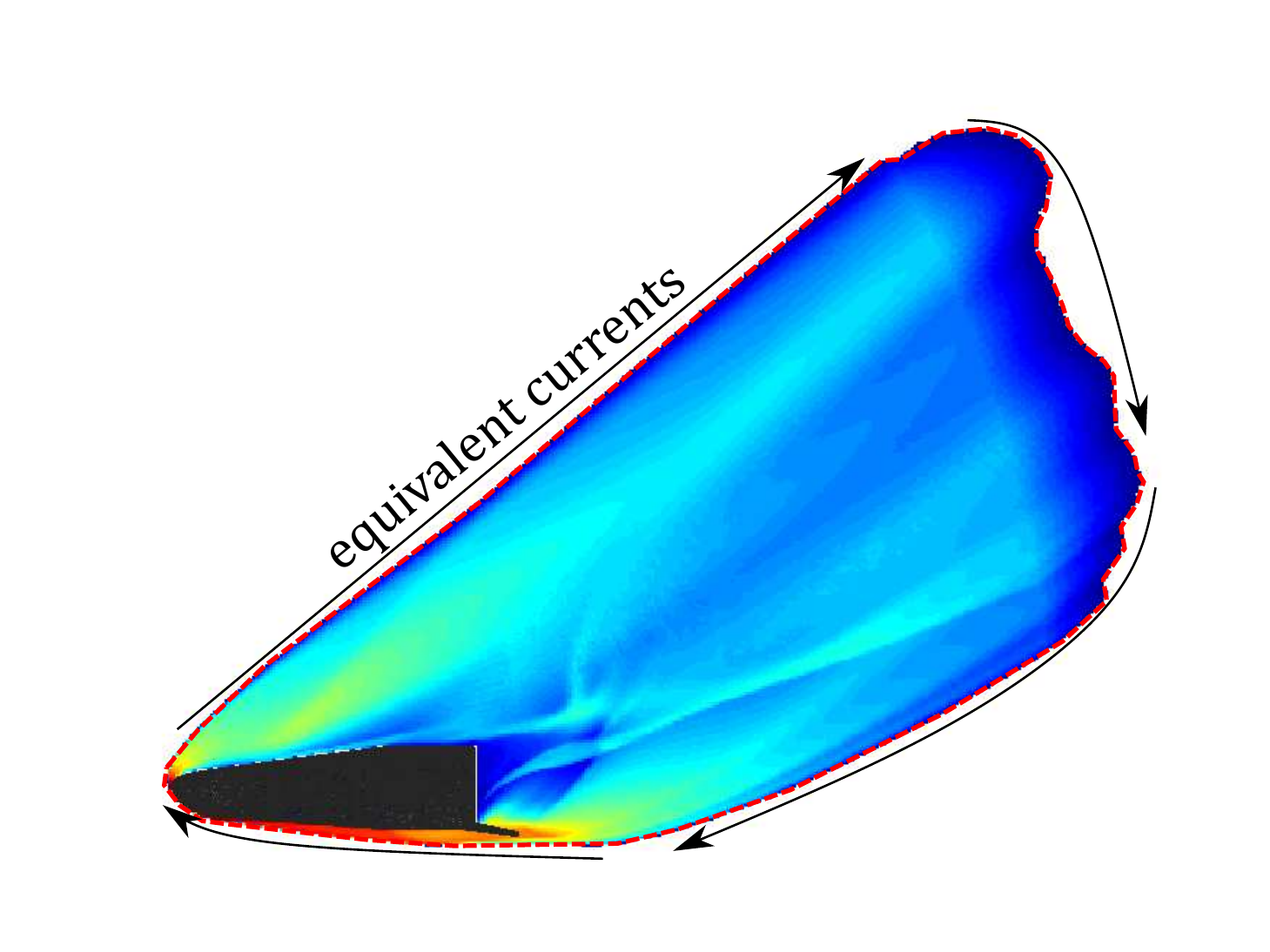}}
		\caption{Schematic of the physical model. Inhomogeneous media (here plasma surrounding a vessel) are enclosed by an equivalent surface (dashed red line) on which equivalent currents radiate to the outside. Note the large extension of the plasma volume.}
		\label{fig:model:physical_model}
	\end{figure}

	\subsection{Modelling Strategy}\label{sec:mod_strategy}
	
	The overall physical model of the scattering problem is shown in Fig. \ref{fig:model:physical_model}. We have one or more impenetrable  objects (in black) immersed into an inhomogeneous medium (coloured map) defined by the dielectric function $\epsilon(\bm{r})$ (possibly complex).
	
	The source can be internal to the inhomogeneous dielectric, as is the case for an antenna on a vehicle, or external, such as an incident plane wave, as it is the case for RCS computations. 
	
	We employ Equivalence Theorem to separate the inhomogeneous region from the surrounding free space via an equivalent (Huygens) surface. The scattered field is obtained via free space radiation  of the magnetic and electric surface equivalent currents $\mathbf{M}_s=-\mathbf{n} \times \mathbf{E}$ and $\mathbf{J}_s=\mathbf{n} \times \mathbf{H}$,  where the $\mathbf{E}$ and $\mathbf{H}$ are the electric and magnetic fields on the equivalence surface, and  $\mathbf{n}$ is the surface normal unit vector. 
	
	We are especially concerned with situations where the plasma volume is electrically large, as is typical in most cases of interest, and its density varies slowly compared to the  operational wavelength $\lambda$; 
	hence, we employ the Eikonal approximation \cite{Born_Optics,Kravtsov_1990,Bremmer_1984} to Maxwell equations in the inhomogenous region; note that the mentioned use of the Equivalence Theorem does not involve any approximation for radiation outside the inhomogeneous region.
	
	Rays originate either from the antenna location (radio link case) or from outside the plasma region (RCS case). 
	There is no physical discontinuity between plasma and free space, avoiding the need to generate both reflected and refracted rays at interfaces. 
	Regardless of their origin point, rays are followed until they intersect the equivalent surface, where their contribution to the (equivalent) field is calculated. For the case of incident wave source, where rays start outside the equivalent surface, both entrance and exit intersections are considered. Once the fields at the equivalent surface are calculated by ray-tracing, they are converted to surface current via the Equivalence Theorem and made radiate in free space generating the far total (radio link case) or scattered (RCS case) field.
	
	It is now appropriate to briefly discuss the choice of the equivalence surface. The Eikonal solution is an approximate one, and thus it is convenient to limit its region of application as much as possible; hence, the equivalence surface is chosen to enclose only points where plasma effects are not negligible (from that point outwards, radiation is exact). This can be implemented in the ray propagation scheme described below in Sec. \ref{sec:model:raypropagation} by stopping the rays where such a condition happens. 
	However, there are situations where rays encounter caustics, and on such surfaces the field approximation is inaccurate; when this happens close to the equivalence surface (see e.g. the discussion in Sec. \ref{sec:benchmark:RCS_inhom}), it is then necessary to move such a surface outwards. 
	
	Finally we observe that in case material discontinuities with penetrable materials need to be modeled, they can be taken into account by using the Fresnel transmission and reflection coefficients, and following both the reflected and refracted rays \cite{Kim92}.
	
	\subsection{Propagation within the vehicle-plasma region}\label{sec:model:raypropagation}  
	
	To compute the EM wave propagation into inhomogeneous media we use the classical Eikonal theory \cite{Kravtsov_1990}. 
	The Eikonal field approximation can be expressed as follows:
	\begin{equation}
		\mathbf{E}(\mathbf{r})=\mathbf{e(\mathbf{r})}E(\mathbf{r})e^{ik_0S(\mathbf{r})}
		\label{eq:Eikonal_approx}
	\end{equation}
	where $E(\mathbf{r})$ is a slowly varying field, $\mathbf{e}$ is the unit polarisation vector, and $S(\mathbf{r})$ is the normalised Eikonal phase function (defining the wave-front surface). This Ansatz is inserted into Maxwell equations, with the assumption that the medium properties are weakly dependent on the spatial coordinate (see below for discussion of validity),
	which yields the hierarchy of equations for the Eikonal function, energy (power density) and the polarization \cite{Kravtsov_1990,Born_Optics}.
	As this is well-known, we will only list our specific choices for the numerical implementation of this set of equations. 
	We express the ray trajectory equation as system of first order Ordinary Differential Equations (ODEs) using the normalised wave-vector $\bm{\xi}=\bm{\nabla} S$ as dynamic variable and the electrical path length differential $d\sigma=\Re(\sqrt{\epsilon_r})~ds$, where $s$ is the arc-length along the trajectory; this results in:
	\begin{eqnarray}\label{eq:ray_eq}
		\frac{d\bm{r}}{d\sigma} &=& \frac{\bm{\xi}}{\Re(\epsilon_r)}
		\label{eq:ray_equation1}\\
		\frac{d\bm{\xi}}{d\sigma} &=& \frac{1}{2\Re(\epsilon_r)}\bm{\nabla} \Re(\epsilon_r)\label{eq:ray_equation2}
	\end{eqnarray}  
	To the next order of the hierarchy, one has the power density and polarization transport, that assume the solution of the trajectory equations.
	The complex polarisation unit vector $\hat{\mathbf{e}}$ obeys the following equation along the ray trajectory \cite{Born_Optics}:
	\begin{equation}
		\frac{d\hat{\mathbf{e}}}{d\sigma} = -\left(\hat{\mathbf{e}} \cdot \frac{d\mathbf{\bm{\xi}}}{d\sigma}\right) \frac{\bm{\xi}}{|\bm{\xi}|^2}
		\label{eq:GOpolarization}
	\end{equation}
	As the trajectory equations (\ref{eq:ray_eq}), this equation is integrated with standard ordinary differential equations (ODE) techniques. At a difference with previous literature \cite{Ling_TAP_1991,Kim92}, we have chosen to use (\ref{eq:ray_eq}) and (\ref{eq:GOpolarization}) because they have shown to be more stable in complex geometries as those present in actual re-entry situations, as in Sec. \ref{sec:IXV_plasma}; this results from the absence of high order curve differentials that would need to be computing \emph{after} ray tracing. Furthermore, these formulas are easily integrated into ODE solution involved in ray tracing, which enables more efficient and accurate computations.
	
	Energy transport requires considering that the plasma is possibly lossy due to collisions; we assume weak to moderate dissipation and follow here the general procedure for anisotropic plasma transport \cite{Bernstein_PF_1975}, for the special case of an isotropic plasma. The transport equation for the field energy can be cast into the form of a divergence equation \cite{Born_Optics} with a linear dissipation term coming from the anti-hermitian part of the dielectric tensor (simply $\Im(\epsilon_r)$ in the isotropic case) as:
	\begin{equation}
		\nabla \cdot ({v} \hat{\mathbf{t}}U)=ck_0\frac{\Im(\epsilon_r)}{\Re(\epsilon_r)}U
		\label{eq:transport}
	\end{equation}
	where $c$ is the vacuum speed of light, $\hat{\mathbf{t}}$ is the unit vector along the trajectory, $k_0$ is the free-space wave number and $U=\Re(\epsilon_r)|E|^2$ is the EM energy density in GO approximation. 
	 Equation (\ref{eq:transport}) shows the interplay between: 1) the divergence of rays due to the local variations of the local index 2) the energy absorption by the lossy medium, embedded in the imaginary part of $\epsilon_r$, and 3) the wave amplitude $|E|$.

Phase and amplitude transport in our case result in the following relationship between (complex) amplitudes at two points $P_1$ and $P_2$ along a ray path:
	\begin{equation}\label{eq:fieldP1P2}
		E(P_2)=E(P_1) \sqrt{\frac{\Re(n_2)}{\Re(n_1)}} DF \exp(i \Phi-\alpha),
	\end{equation}
	where $n_{\ell}=\sqrt{\epsilon_r(P_{\ell})}$ is the complex refractive index evaluated at $P_{\ell}$, with $\ell = \{1, 2\}$, and
	\begin{equation}\label{eq:fieldP1P2_terms}
		\alpha= k_0/2 \int_{\gamma}\frac{\Im\left(\epsilon_r(\bm{r})\right)}{\Re\left(\epsilon_r(\bm{r})\right)}d\sigma, \ 
		\Phi=k_0 \int_{\gamma} d\sigma,
	\end{equation}
	where $\gamma$ is a parametrization of the ray trajectory between points $P_1$, and $P_2$, namely $\gamma: [\sigma_1, \sigma_2] \rightarrow \mathbb{R}^3$ with $\gamma(\sigma_1) = P_1$ and $\gamma(\sigma_2) = P_2$ and $\gamma$ obeys to \eqref{eq:ray_equation1} and \eqref{eq:ray_equation2}.
	In \eqref{eq:fieldP1P2} $DF$ indicates the divergence factor that accounts for divergence of the infinitesimal ray tube (i.e. along an individual ray) in a lossless medium; it represents the shrinking or expansion of the wave front surface along its propagation, and will be addressed next in Sec.\ref{sec:model:raytube}. The medium properties needed for evaluating the ray path and equations (\ref{eq:fieldP1P2},\ref{eq:fieldP1P2_terms}) are either computed from analytical models or via numerical interpolation of gridded data as explained in Sec. \ref{sec:plasma_model}.
	
	\subsubsection{Discussion on GO validity for hypersonic plasmas}\label{sec:model:validity}	
	
	Following well established literature \cite{Kravtsov_1990,Born_Optics,Bernstein_PF_1975} the necessary condition for the applicability of the ray theory can be cast into the following relations:
	\begin{eqnarray}
	\delta \equiv \frac{|\nabla\log(|\epsilon_r|)|}{k_0} &\ll& 1
	\label{eq:validity_1} \\
	 \frac{|\nabla\log(|\mathbf{E}|)|}{k_0} &\ll& 1
	 \label{eq:validity_2} \\
	\frac{\Im(\epsilon_r)}{\Re(\epsilon_r)}  &\sim& \delta 
	\label{eq:validity_3}
	\end{eqnarray}
	
	The first two demand the properties of the medium and the wave amplitude to vary slowly over the distance of a wavelength (i.e. weakly dependent on the spatial coordinate $\mathbf{r}$). The last equation requires a weak energy exchange between the media and the EM wave and partly overlaps with eq. \ref{eq:validity_2}. These conditions are typically violated near caustics (even in free-space), cut-off and resonance regions. Here full wave solution is needed. 
	Also, for a general lossy media violating eq. \ref{eq:validity_3},  the conventional definition of the group velocity leads to complex velocities even at leading order in $\delta$. Consequently, the rays lay, in general, in a complex space-time. The difficulties arise, in this case, from the interpretation of these complex rays and their connection with the 'real' world. Several models have been proposed all leading to a modified ray trajectory due to the strong absorption. The interested readers may look for instance at the following references \cite{Jones_RS_1970,Censor_CEEE_1989}.
The case study presented in the section \ref{sec:benchmark:RCS_inhom} are chosen such that the conditions of validity are well satisfied (see section \ref{sec:benchmark:RCS_inhom}). In a hypersonic plasma the situation is more complicated. A high density, high gradient region may be generated near the hypersonic vessel and depending on the flight conditions (i.e. Mach number) cut-off layers appear. The ray is expected to suffer reflection at cut-off, since the wave is evanescent on the other side. Actually, reflection must occur before the cut-off is reached \cite{Brambilla_PoP_1982_RT_LH}. Indeed, for oblique propagation, only the component of $\mathbf{k}$ parallel to the density gradient goes to zero, whereas the group velocity is still well defined. This kind of reflection does not lead to a global break-down of the Eikonal approximation, but is associated with the formation of a caustic or of a focal point, so that some care has to be taken in the determination of the phase along the ray (see section \ref{sec:model:rayintegration:fold}). Thus, even in this extreme case, a good approximation is expected away from the cut-off layer (i.e. at the equivalent surface).
In section \ref{sec:IXV_plasma} we discuss the validity of the ray approach for each considered plasma case.

\subsection{Ray-Tube approach}\label{sec:model:raytube}
	
\begin{figure}[!ht] 
		\centerline{\includegraphics[width=0.70\columnwidth]{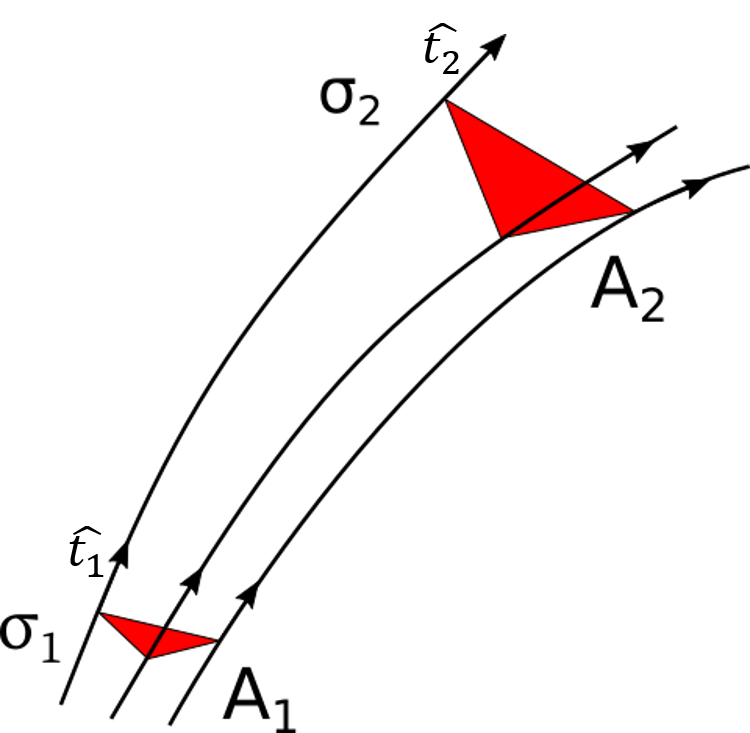}}
		\caption{Sketch of the adopted ray tube propagation scheme.}
		\label{fig:model:tube_propagation}
	\end{figure}

\begin{figure}[!ht] 
		\centerline{\includegraphics[width=0.99\columnwidth]{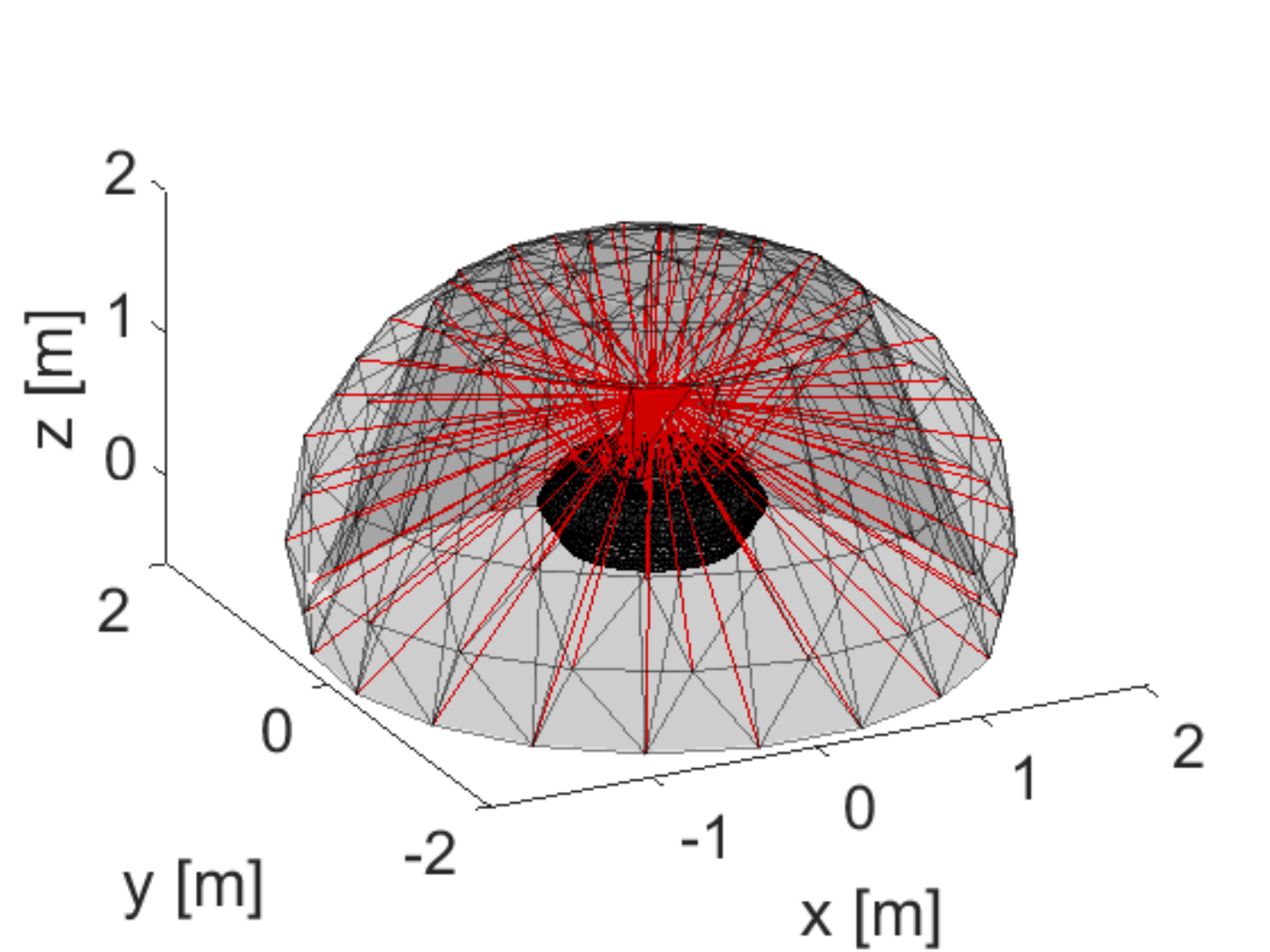}}
		\caption{Ray trajectories (red lines) from a dipole source above a small simplified vessel, in free space. The spherical equivalence surface is also depicted (coloured).}
		\label{fig:model:rays_trajectory}
	\end{figure}

	Because of the expected complexity of the ray set as a whole, we shift from a ray-based approach, as common in the related literature, to a ray-tube based description of the relevant field quantities. This allows to include collective effects such as ray folds (e.g. caustics) that are not directly present in the individual ray descriptors, and to express amplitude and radiation in a more effective manner.
	To do this, we group rays in bundles of three (see figures \ref{fig:model:tube_propagation} and \ref{fig:model:rays_trajectory}), on the basis of geometrical contiguity at launch; this only entails proper book-keeping of trajectories. Each three-ray bundle is a ray-tube in our scheme.
	
\begin{figure}[!ht] 
		\centerline{\includegraphics[width=0.99\columnwidth]{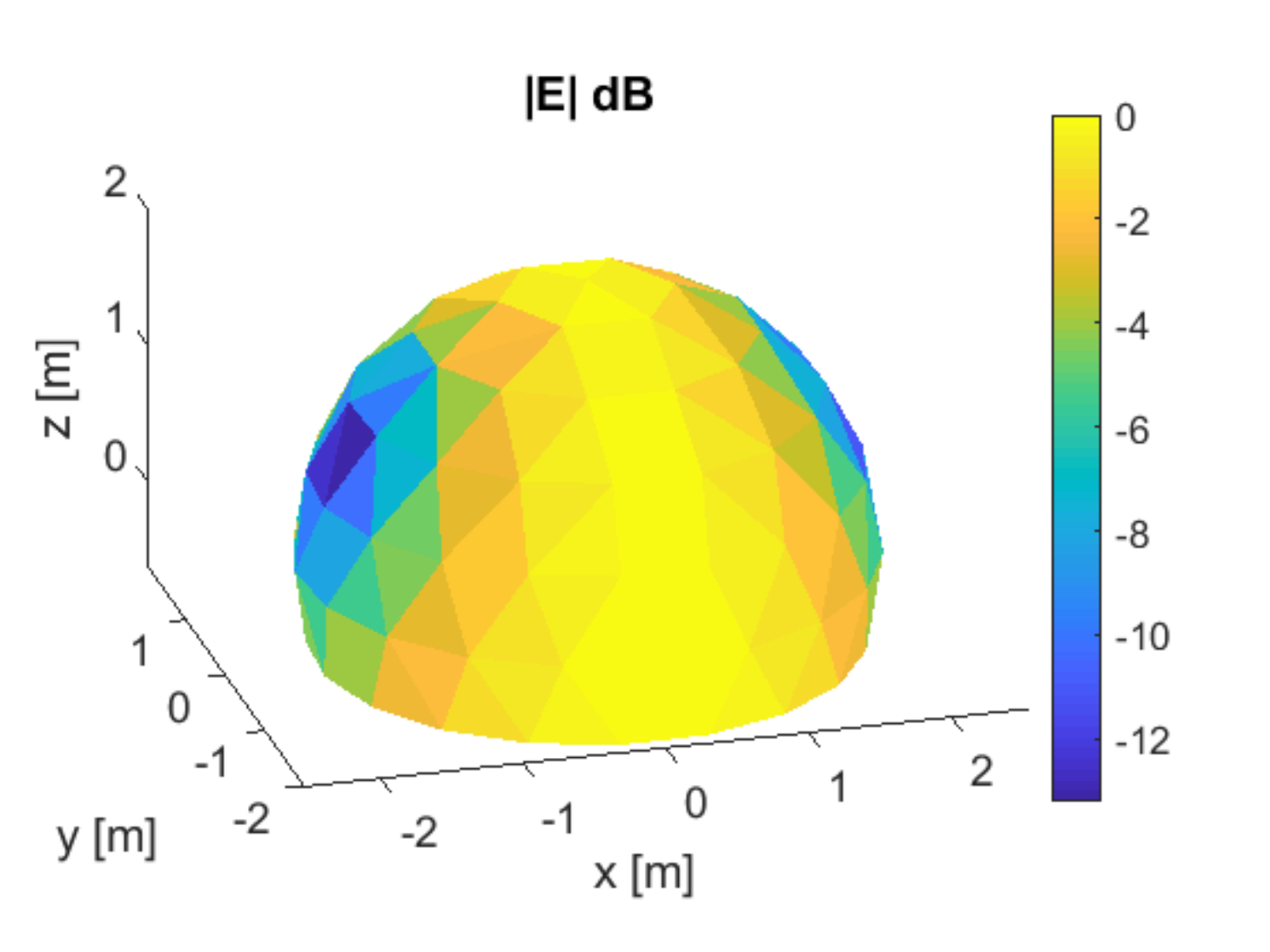}}
		\caption{Electric field on a spherical equivalent surface for the configuration in Fig. \ref{fig:model:rays_trajectory}.}
		\label{fig:model:E_on_equiv_surf}
	\end{figure}

We begin by briefly describing the ray initialisation scheme. As a first step, the equivalence surface (ES) is converted into a triangular tessellation using standard algorithms; the triangular cells have a typical side length varying in the range from $\lambda/10$ to $\lambda/3$.
As said earlier on, the equivalence surface is arbitrary and may coincide with the boundary of the scattering media. Alternatively, a canonical shape such as a sphere (see fig. \ref{fig:model:rays_trajectory} and \ref{fig:model:E_on_equiv_surf}) can be chosen to encapsulate the media volume. 
The nodes (vertices) of the mesh determine ray launch directions. For sources in the plasma region (e.g. antennas mounted on a vehicle) ray starting directions are obtained by the directions of the vertices of the equivalent surface mesh as seen from the source location. For complex antennas, this is clearly a simplistic description for two reasons: a) a unique source (or phase center) may not be enough to accurately describe the near field and b) the interaction of this of this latter with the plasma background is neglected. We argue, however, that for antennas considered in practical applications (i.e TTC or NAV type) are such that a single source point is enough. This can be extended with multiple source points for large antennas, and we actually did it when analysing arrays (not reported here). Moreover, if there is a significant near-field interaction with the plasma, one needs a full-wave simulation for that. For a single antenna (as opposed to arrays) that may significantly impact on input impedance when the interaction is strong, much less so for radiation pattern. That near field analysis can be done in several ways, and actually one of the authors has extensive experience with antennas facing complex (magnetized) plasmas \cite{Lancellotti_NF2006_TOPICA}. However, we deemed that issue was off track in this case, which focuses on the plasma effect on radiation.  For a plane wave source, rays start all parallel from a plane just outside the equivalence surface, with initial direction normal to the starting plane; their location (and density) is determined by back-projecting the mesh nodes along the wave direction onto the starting plane.
	
	The initial associated field intensity and polarisation may be sampled from input antenna data or may be provided in analytical form.  In figure \ref{fig:model:rays_trajectory} we show a set of rays (red lines) launched from a point just above a simple vessel (black), in free space. The rays point toward the spherical equivalence surface nodes and they are stopped at the intersection. 
	
	Rays that encounter the impenetrable vessel (or any other such body) are reflected, i.e. they are stopped and re-launched with new initial conditions according to the usual specular reflection laws and the appropriate Fresnel reflection coefficients.
	Figure \ref{fig:model:E_on_equiv_surf} shows an example of the electric field on the equivalence surface. 
	

	\subsubsection{Amplitude and Radiation} \label{sec:model:rayintegration}
	Convenience of the ray-tube description is immediately evident for the amplitude calculation, and the ensuing radiation. The divergence factor (DF) appearing in amplitude propagation (\ref{eq:fieldP1P2}) can in principle be expressed in terms of the local surface principal radii (see for example \cite {Bremmer_1984}). Although mathematically exact, this formulation contains higher order derivatives of the Eikonal function which may be subject to large error when the medium properties are not known analytically but resulting from other numerical calculations or measurements, as in our CFD-originating plasmas. We prefer instead to estimate DF directly from the wave front evolution of the three-ray tubes introduced above
	(see figures \ref{fig:model:tube_propagation} and \ref{fig:model:rays_trajectory}). With reference to the notation of Fig.~\ref{fig:model:tube_propagation}, we have that the divergence factor can be related to the changes in the front surface area simply with the relation $DF = \sqrt{A_1/A_2}$.
	
	
	\begin{figure}[!ht] 
		\centerline{\includegraphics[width=0.99\columnwidth]{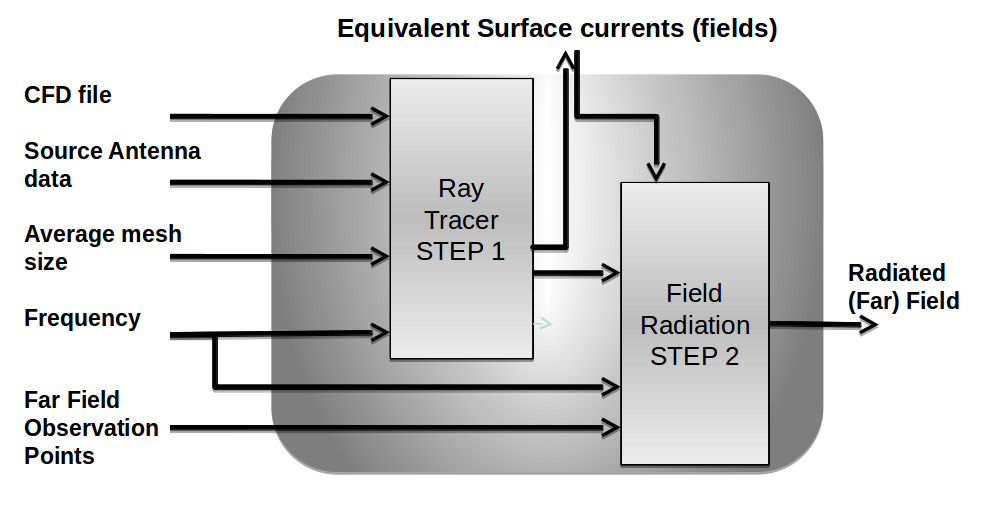}}
		\caption{Algorithm overview with input and output data.}
		\label{fig:model:algorithm_overview}
	\end{figure}
	
	
	At a difference with other approaches \cite{Kim92,Ling_TAP_1991}, the radiated field 
	is obtained consistently with the ray-tube based approach described above. That means that the (far) radiated field is expressed as the sum of the radiation from the patches $\Delta T_i$ defined by the intersection of the ray {\em tubes} with the chosen equivalence surface, 
	\begin{equation}
		\mathbf{E}_s (r,\theta, \phi)= \sum_i \mathbf{E}_{s,i} (r,\theta, \phi)
		\label{eq:rad_sum}
	\end{equation}
	an example of these patches is shown in the simple case of Fig. \ref{fig:model:E_on_equiv_surf}.
	

	The field on each triangular patch is approximated with constant magnitude and linear phase variation. The electric field vector $\mathbf{E}_i$ is obtained as the average of the three values pertinent to the three rays of any tube, as per (\ref{eq:fieldP1P2}), (\ref{eq:GOpolarization}) (note that the divergence factor is intrinsically a tube property, while phase, attenuation and polarization are properties of individual rays). The magnetic field $\mathbf{H}_i$ is likewise obtained from the three values at the rays, in turn related to the electric field there by the local ray impedance relationship, and the wavefront normal $\hat{\bm{t}_i}$ is likewise the average of the three ray wave front normals. 
	
	With these assumptions, the patch radiated fields can be expressed as:
	\begin{equation}
		\mathbf{E}_{si} (r,\theta,\phi) = \frac{-ik_0}{4\pi} \frac{e^{ik_0r}}{r} e^{-i k_0\hat{\mathbf{r}} \cdot \mathbf{r}_i} S_i(\theta,\phi) \mathbf{a}_i \\
		\label{eq:rad_int}
	\end{equation}
	with 
	\begin{eqnarray}
		\begin{aligned}
			& \mathbf{a}_i= \hat{\mathbf{\theta}} (-\hat{\mathbf{\phi}} \cdot \mathbf{M}_i + \hat{\mathbf{\theta}} \cdot Z_0 \mathbf{J}_i) + \hat{\mathbf{\phi}} (\hat{\mathbf{\theta}} \cdot \mathbf{M}_i + \hat{\mathbf{\phi}} \cdot Z_0 \mathbf{J}_i) \\
			& S_i(\theta,\phi)= \frac{1}{\Delta T_i}\int_{\Delta T_i} e^{i k_0\left(\hat{\mathbf{r}}-\hat{\mathbf{t}}\right) \cdot \left(\mathbf{r'}-\mathbf{r}_i \right)} d\Sigma'\\
			& \mathbf{M}_i=-\hat{\mathbf{n}}_i \times \mathbf{E}_i, \ \mathbf{J}_i=\hat{\mathbf{n}}_i \times \mathbf{H}_i
		\end{aligned}
		\label{eq:rad_int_patch}
	\end{eqnarray}
	where $\mathbf{r}_i$, $\hat{\mathbf{n}}_i$ and $\Delta T_i$  are the patch centroid, normal unit vector and  the surface area respectively, $\hat{\mathbf{\theta}}$, $\hat{\mathbf{\phi}}$ are the unit vectors in the directions $\theta$, $\phi$ and $\hat{\mathbf{r}}(\theta,\phi)$ the unit vector which points from the origin to the observation point, and $Z_0$ is the free space impedance.
	It turns out that the radiation shape factor $S_i$ can be conveniently written in closed form \cite{Gibson2008} for quick and efficient radiation calculation.
	
	We observe that this ray-tube based approach to radiation does not require to assign a surface area to individual rays on the equivalent surface. Also, in presence of complex inhomogeneity, tubes with rays diverging far apart are expected to have lower accuracy; as they result in larger areas, their contribution to radiated field is weaker: this results in an intrinsic robustness.

	
	Equation (\ref{eq:rad_int_patch}) provides the numerical model for the radiated field, and together with the ray-tracing model in section \ref{sec:model:raypropagation}, constitute the main components of the method proposed.  In figure \ref{fig:model:algorithm_overview} we present a schematic overview of the implemented algorithm, consisting in two main modules: the ray-tracer module and the radiation model together with the inputs and outputs.
	

	\subsubsection{Folds}\label{sec:model:rayintegration:fold}
	It is well known that the Eikonal approximation breaks down near caustics where the field becomes singular. We are, in general, not interested in the field nearby the caustics; thus, we will place the equivalent surface sufficiently far from them (although one must often compromise with other numerical constraints, see section \ref{sec:benchmark:RCS}). Nevertheless, caution must be taken when a ray tube passes throughout one these singular regions. Caustics arise when the ray field folds. The wave front folding affects the phase of the field  carried by the wave with a phase shift of $\pi/2$ radians. These events must be accounted for, so that the proper phase shifts can be applied to the ray tube field. 
	Caustics may be, in principle, detected by calculating the principal radii of the wave front \cite{Kim92,Bremmer_1984}, but again high order numerical derivatives can make this a very difficult task. We detect instead the crossing of rays forming the ray tube which 
	occurs when passing through the caustic point and apply phase shift correction. \\

	\section{Benchmark of numerical model}\label{sec:benchmark}
	
	Validation of the code is divided into two parts: a) against an inhomogeneous dielectric with impenetrable core with spherical symmetry; b) for complex cores in free space.
	Study case a) mimics the configurations of interest, a flying object enveloped in a non-homogeneous plasma, but in a geometry that affords exact results for reference. Because of availability of reference results, the spherical validation a) will be done with plane wave excitation. Conversely,  free space validation b) can be done against reference Method of Moments (MoM) simulations for both simple and real-life vehicle shapes; the latter will be done especially for testing on-vessel near sources: We employ \cite{Li_TAP_2015} to generate the MoM reference.
	
	In the following, the present hybrid method will be often labelled ''RT'' (after Ray Tracing) for the sake of brevity.
	
	
	

	
	\subsection{Validation against spherical symmetry references }
	\label{sec:benchmark:RCS}
	
	Thanks to the existence of accurate analytic solutions using the Mie series \cite{1970_Ruck_RCS, Balanis_EM, Mie_WG} for spheres we can compare our results with highly accurate references.
	
	The electrical sizes considered in the following are on the same order of magnitude of those involved in the real-life application case in Sec. \ref{sec:IXV_plasma}; as seen in Fig.\ref{fig:model:IXV}, the IXV vehicle is enclosed in a box of size $ L_1 \times L_2 \times L_3$, $ L_1=5$m, $ L_2=2.2$m, $ L_3=1.7$m; at the S-band telemetry and command link frequency of 2.6GHz the electrical size of the box is $ k L_1 \approx 300 $, $ k L_2 \approx 120 $, $ k L_3 \approx 90 $, where $k=2 \pi /\lambda$ is the wave number. 
	
	\subsubsection{PEC sphere in vacuum}
\begin{figure}[!ht] 
		\centerline{\includegraphics[width=0.99\columnwidth]{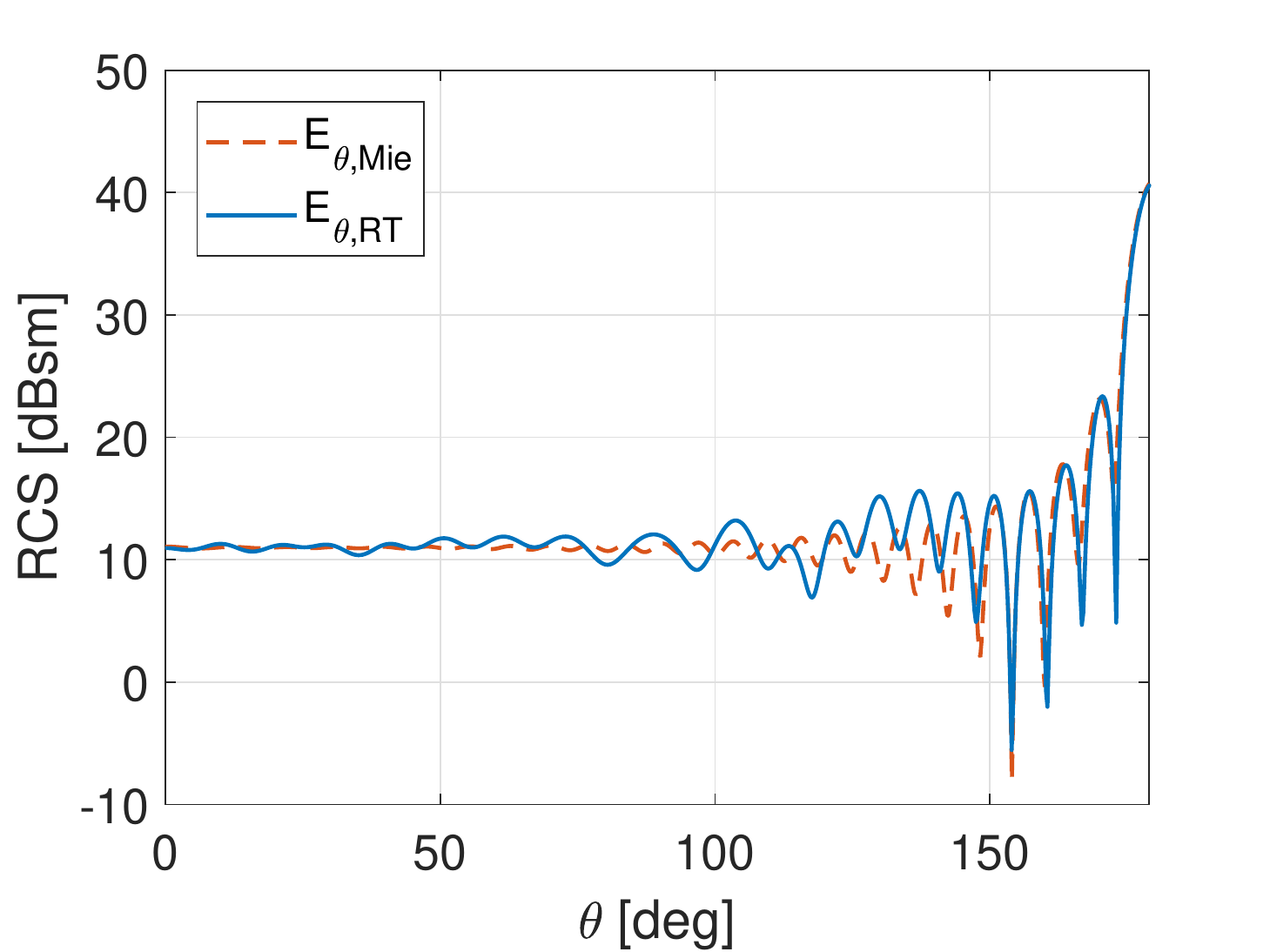}}
		\caption{Bistatic RCS of a PEC sphere for $ka=31.44$. Comparison of present method (labelled RT) and Mie solutions.}
		\label{fig:benchmark:PEC_sphere}
	\end{figure}

\begin{figure}[!ht] 
		\centerline{\includegraphics[width=0.99\columnwidth]{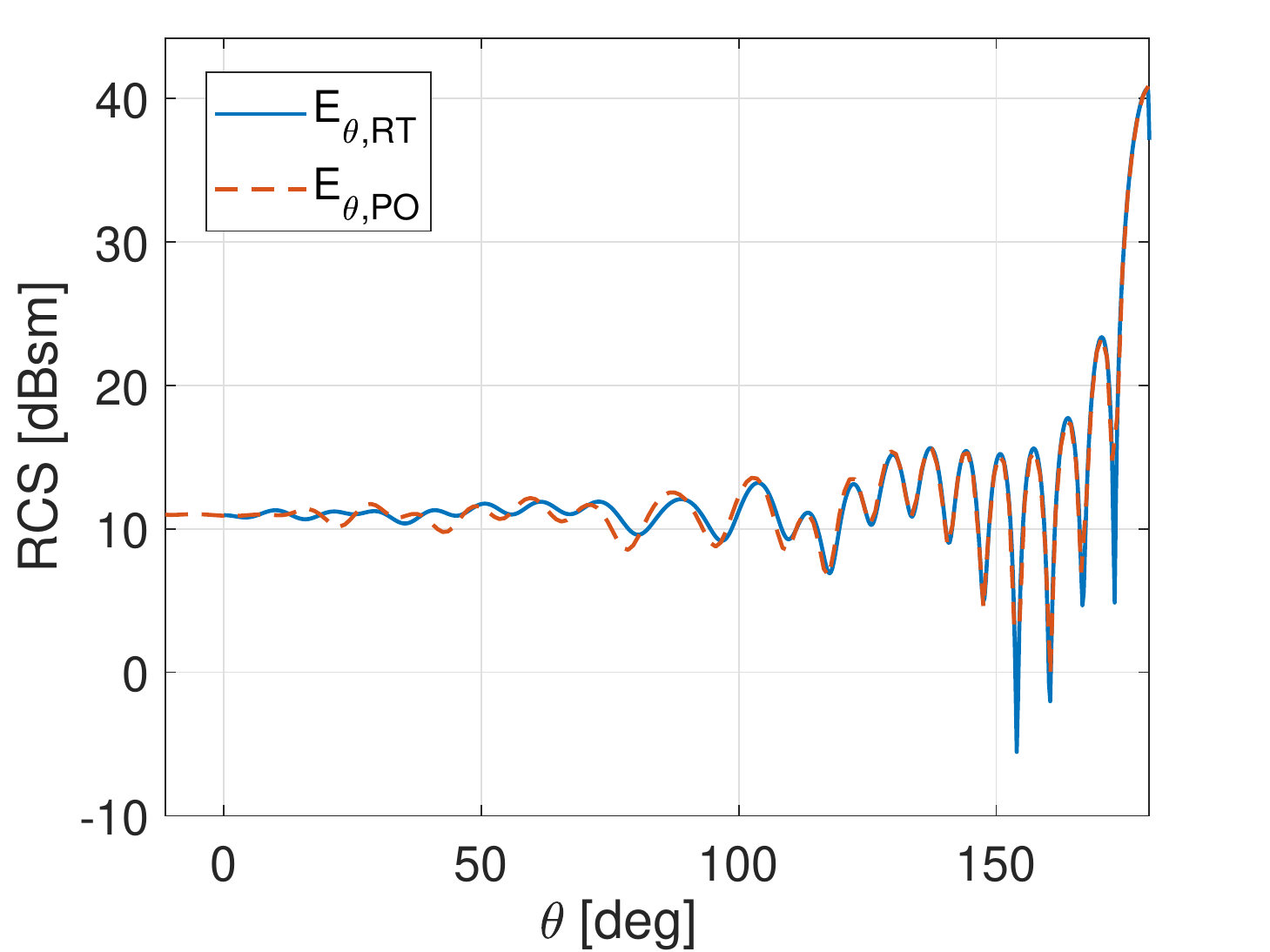}}
		\caption{Bistatic RCS of a PEC sphere for ka=31.44. Comparison of present method (labelled RT) and PO solutions}.
		\label{fig:benchmark:PEC_sphere_PO}
	\end{figure}

\begin{figure}[!ht] 
		\centerline{\includegraphics[width=0.99\columnwidth]{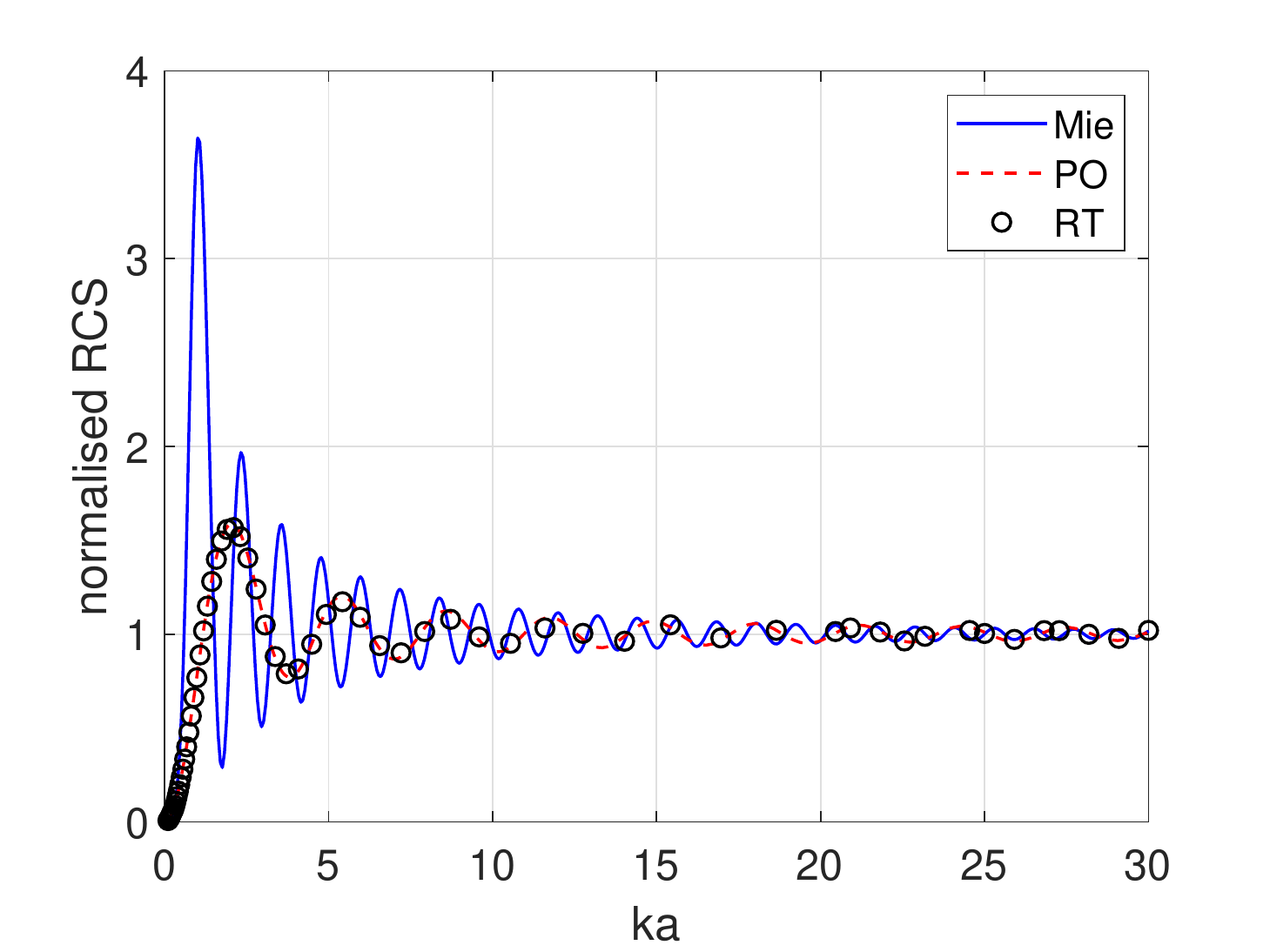}}
		\caption{RCS of a PEC sphere as function of ka. Comparison of present method (labelled RT), PO, and Mie solutions.}
		\label{fig:benchmark:RCS_ka}
	\end{figure}

	The only physical parameter which varies in the case of a PEC sphere in free space \cite{Wait1965, Balanis_EM} is the product $ka$, where $a$ is the sphere radius. As for all asymptotic methods, good results are expected for large $ka$. Forward and back scattering values computed with the present method are found to agree within 10$\%$ of the Mie series values when $ka>$10 (back scattering shown in Fig~\ref{fig:benchmark:RCS_ka}). Figure \ref{fig:benchmark:PEC_sphere} shows the bistatic RCS the vertical or  $\theta$-polarization with $ka=$31.4  
calculated with the Mie series and the present method. The horizontal polarisation is omitted for clarity and presents similar features. Here the convention is that the plane wave (PW) propagates along $-z$, thus we have back-scattering for $\theta$=0$^\circ$ and forward-scattering for $\theta$=180$^\circ$. The diffractive interference pattern in the angular region of forward scattering (large angles) is qualitatively reproduced with decreasing accuracy toward smaller angles where oscillations between Mie and our solutions become out of phase. 
	We notice that the results of the present method are very similar to those of Physical Optics (PO)  (Fig.~\ref{fig:benchmark:PEC_sphere_PO} and ~\ref{fig:benchmark:RCS_ka}): we now explain this result. As said earlier, the location of the equivalent surface is arbitrary but best results for this case are obtained by placing the surface as close as possible to the PEC. Under these conditions, the resulting equivalent currents are very close to the PO currents $\mathbf{J}=2\mathbf{H}_{inc} \times \mathbf{n}$ and $\mathbf{M}=0$ where $\mathbf{H}_{inc}$ is the incident magnetic field and $\mathbf{n}$ is the normal to the PEC surface. For ray fields, the total field is the sum of the incident term plus the specular reflected field. When the equivalent surface approaches PEC surface the tangent electric field $(\mathbf{E}_{inc} + \mathbf{E}_{ref})\times \mathbf{n}$ cancels out, and the magnetic field doubles; this fact has been verified numerically using equivalent surfaces at distance $\leq$0.01$\lambda$ from the PEC. 
	Hence, 
	the results of the proposed hybrid method are consistent with the result of  PO, as this latter is actually  the limit for this hybrid method in these configurations. 
	
	
	\subsubsection{Inhomogeneous sphere with PEC core} 	\label{sec:benchmark:RCS_inhom}
\begin{figure}[!ht] 
		\centerline{\includegraphics[width=0.99\columnwidth]{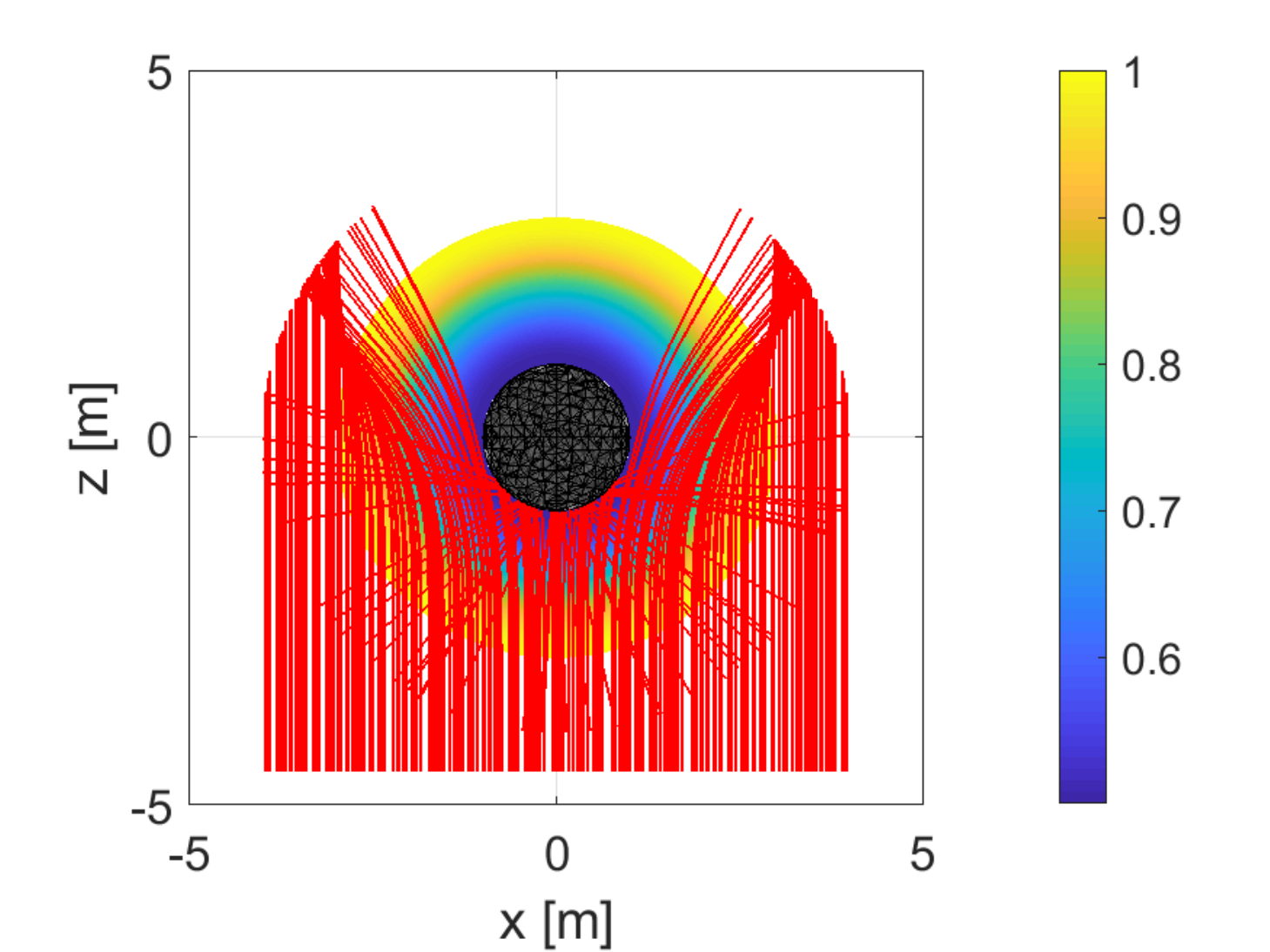}}
		\caption{Spherically inhomogeneous dielectric with PEC core,  $\epsilon_r(r)<1$: selected ray trajectories (line source). PEC sphere is in black, the colormap represents $\epsilon_r(r)$  with the profile in eq. (\ref{eq:Ling_profile}) and parameters $\epsilon_c=$0.5, $R_{out}/R_{in}=3$, and $kR_{in}=62$.}
		\label{fig:benchmark:PEC_plasma_sphere_rays}
	\end{figure}

	We now consider an inhomogeneous dielectric surrounding a PEC core, both with spherical symmetry. Despite the simplicity of the spherical geometry, this example includes all the complex physical phenomena and conditions of a real-life case such as ray bending (due to continuous refraction), reflection and diffraction. Indeed, as will be seen, this is a conservative test, as the (extreme) symmetry entails special critical cases  that are unlikely in more complex geometries. The  dielectric has a permittivity profile given by: 
	\begin{equation}
		\epsilon_r(r)=\frac{1+\epsilon_c}{2} \left(1-\frac{1-\epsilon_c}{1+\epsilon_c} \cos{\left( \pi\frac{r-R_{in}}{R_{out}-R_{in}}\right) }\right) 
		\label{eq:Ling_profile}
	\end{equation}
	with  $R_{in}=a$, $R_{out}=3 R_{in}$; we will examine scattering as a function of the relative permittivity at the core, $\epsilon_c$. As before, we focus on electrically large objects and consider a PEC sphere with radius $ka$=62 (e.g. $a=$1m at a frequency of $f=$3GHz). The scattering geometry and selected ray trajectories are shown in figure \ref{fig:benchmark:PEC_plasma_sphere_rays}. 
	
	\begin{figure}[!ht] 
		\centerline{\includegraphics[width=0.99\columnwidth]{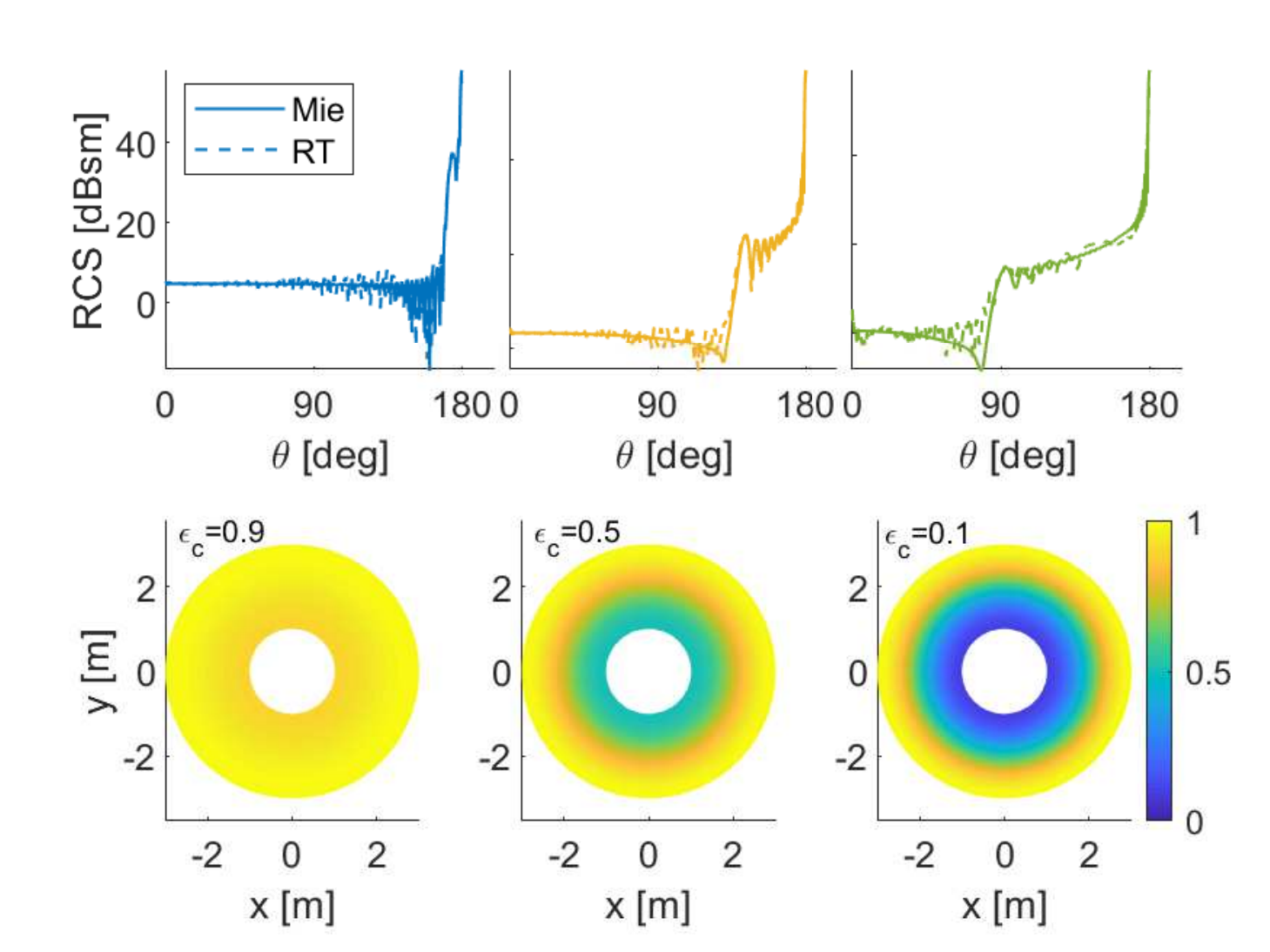}}
		\caption{Spherically inhomogeneous dielectric with PEC core,  $\epsilon_r(r)<1$: bistatic RCS for various values of minimum permittivity ($\epsilon_c$). The colormap represents $\epsilon_r(r)$  with the profile in eq. (\ref{eq:Ling_profile}) with $R_{out}/R_{in}=3$, $kR_{in}=62$, and for various listed $\epsilon_c$.
			Upper row: RT Vs. Mie results for $\epsilon_c=$0.9-0.1. Lower row: color map of the relative permittivity profiles.}
		\label{fig:benchmark:PEC_plasma_sphere_RT_Mie}
	\end{figure}
	
	This dielectric profile has the effect of bending the rays away from the sphere, producing field scattering and a peaked bistatic RCS around $\theta=0^\circ$. 
	
	Figure \ref{fig:benchmark:PEC_plasma_sphere_RT_Mie} displays the bistatic RCS profiles for $\epsilon_c$ in the range $0.1-0.9$ for both the RT and Mie solution. Even at the lowest permittivity $\epsilon_c \leq 0.1$ the small parameters $\delta$ (see eq. \ref{eq:validity_1}) is mostly below 0.01 and never exceeds 0.05 for all traced rays, ensuring the validity of the ray tracing method.
Here the convention is that the PW propagates along $-z$, thus having back-scattering for $\theta=0^\circ$ and forward-scattering for $\theta=180^\circ$.	
	
\begin{figure}[!ht] 
		
		\centerline{\includegraphics[width=0.99\columnwidth]{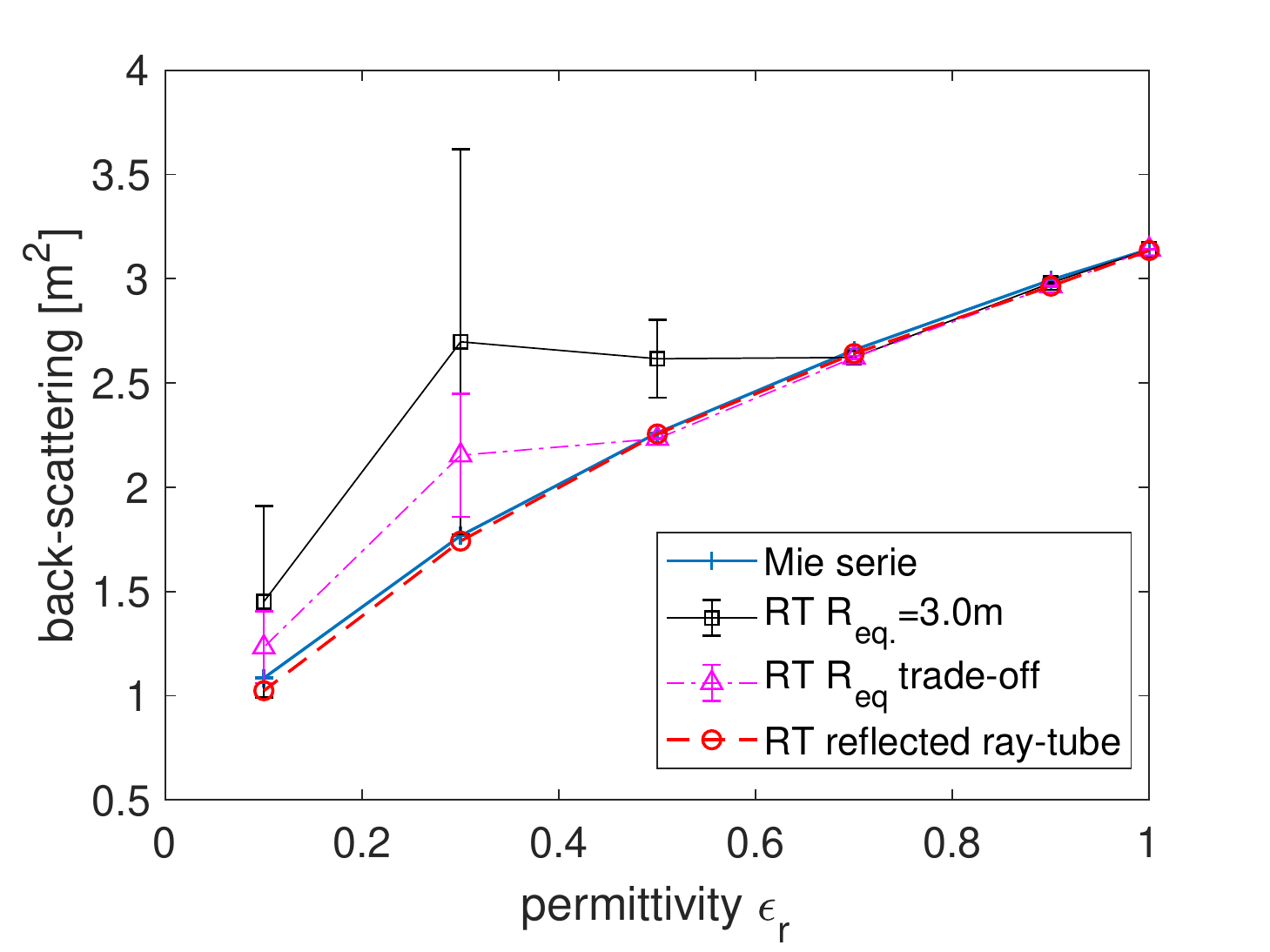}}
		\caption{Spherically inhomogeneous dielectric with PEC core,  $\epsilon_r(r)<1$: Comparison of the monostatic RCS for $\epsilon_c$ ranging from 1 to 0.1. Effect of choice of Huygens surface and ray-tube selection. Default Huygens at dielectric boundary with $R_{Huygens}=3.0m$ in black, trade-off $R_{Huygens}$ in magenta and extraction of reflected ray-tube only in red.}
		\label{fig:benchmark:Plasma_PEC_back_scattering}
	\end{figure}

	There are three major qualitative changes when varying $\epsilon_c$: 1) the back-scattering decreases for lower $\epsilon_c$; 2) the peak forward scattering ($\theta=180^\circ$) increases with scattering volume but does not depend on $\epsilon_c$ (note that the PEC sphere in vacuum has RCS$_{vac}$=47 dBsm whilst with dielectric RCS$_{die}$=58 dBsm); 3) the forward scattering profile broadens with lower $\epsilon_c$. The same trends are also clearly visible in the RT simulations (dashed lines) and they find an easy interpretation in terms of rays. At lower $\epsilon_r$ in fact, the rays are more and more refracted away and some ``miss'' the PEC object (see Fig.~\ref{fig:benchmark:PEC_plasma_sphere_rays}) reducing back-scattering and broadening the forward scattering profile. The same mechanism is also at the basis of the plasma stealth effect for not absorbing plasmas (refraction only). Special attention and consideration is needed for solutions with lower $\epsilon_c$. Although qualitatively satisfactory, the profiles with $\epsilon_c<$0.5 suffer from noise and in general lower accuracy so that for those profiles in Fig.~\ref{fig:benchmark:PEC_plasma_sphere_RT_Mie} we have applied some smoothing to reduce small scale spatial features (with dimension smaller than $\lambda$) in order to increase clarity. We have investigated the origin of this inaccuracy. The different ray tube contributions can be divided into: a) tubes entering into the equivalence surface, b) bent and eventually reflected exiting tubes. The analysis of these classes shows that the spurious oscillations can be attributed to the bending rays and ultimately to numerical error in the trajectories, \emph{and} to the appearance of caustics. Caustics surfaces form near and within the dielectric boundary when $\epsilon_c$ falls sufficiently low. In Fig.~\ref{fig:benchmark:PEC_plasma_sphere_rays} the formation of a caustic surface can be observed just outside the outer boundary of the dielectric for $\epsilon_c=0.5$. For an equivalent surface nearby caustics ray field amplitude is small and its evaluation inaccurate; one has then to move this surface away from the object, to have higher field accuracy. This is however sub-optimal, as the best results of the hybrid RT-surface-radiation approach are obtained by placing the equivalent surface as close as possible to the scatterer, reducing the volume where the ray approximation is applied, as discussed earlier on. For this geometry and permettivity distribution, we find a good compromise by positioning of the equivalence surface between 5$\lambda$ and 10$\lambda$ away from the dielectric boundary. This $\lq$optimal' or trade-off surface position changes obviously with the caustics location and ultimately with permittivity. This effect is displayed in Fig.~\ref{fig:benchmark:Plasma_PEC_back_scattering} where we compare the monostatic RCS from Mie series and RT. It is seen that, for $\epsilon_c\leq0.5$, standard Huygens surface at the dielectric boundary leads to large uncertainty in the RCS value, error that is reduced by trade-off placing. Residual error remains for the lowest $\epsilon_c$. At a first glance one may conclude that extraction of back scattering signals is very inaccurate under these conditions. However, the solution is within the framework of the ray approximation itself by isolating the reflected-only ray tubes and their contributions to the back-scattering (see red-dotted line in Fig.~\ref{fig:benchmark:Plasma_PEC_back_scattering}).
	We conclude by stressing that the caustics issue discussed in this example is very extreme and due to the idealised spherical geometry. In real life situations ray trajectories are more chaotic, caustics are not structured, and much less of a problem.  
	
	\subsection{Complex cores in vacuum: comparison with MoM }
	
	\subsubsection{Radiation from on-vehicle antennas}	
	
	
	In this section we benchmark the RT code with simple antennas, although the code can take any antenna pattern as input for the ray initialisation. 
	
\begin{figure}[!ht] 
		\centerline{\includegraphics[width=0.99\columnwidth]{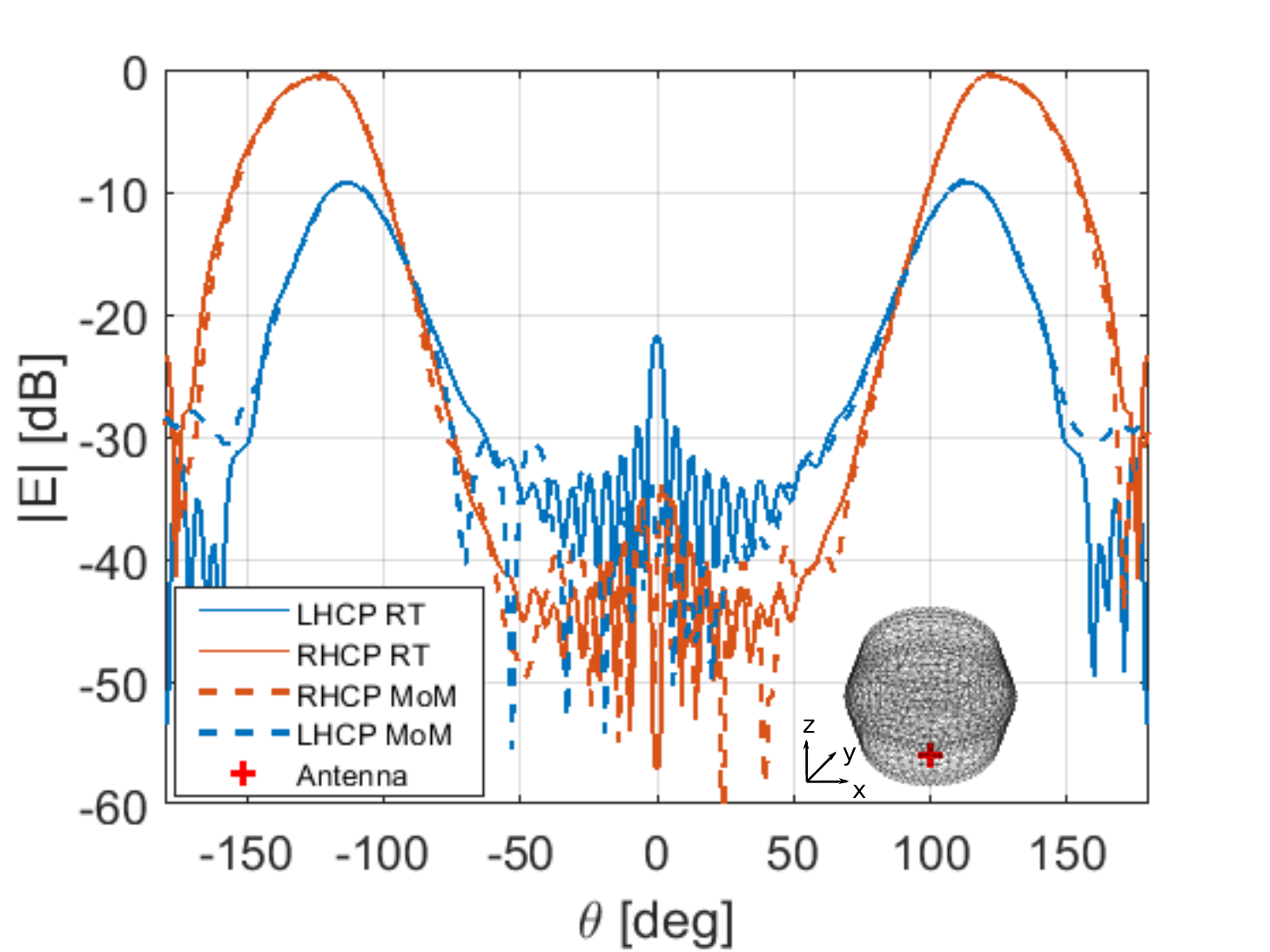}}
		\caption{Radiation of a circularly polarized source on PEC simplified core. The antenna is a pair of crossed dipoles (along x and y), with RHCP. Cut at $\phi=0$.}
		\label{fig:benchmark:MR_comp}
	\end{figure}

	Curved reflecting surfaces demand for sufficiently accurate meshing; hence, before addressing a real-life geometry, we consider a simplified vessel  with surfaces both flat and with single curvature.  The object is shown in the lower right of Fig.~\ref{fig:benchmark:MR_comp}; it is composed of two truncated cones joined together. The top and bottom flat disks are 10$\lambda$ of diameter, and the antenna is located at about 0.25$\lambda$ below the bottom side. Here the antenna is constituted by 2 crossed dipoles fed to provide circular polarization.
	
	The benchmark of the field pattern is presented in figure \ref{fig:benchmark:MR_comp}. Both polar and co-polar polarization are well calculated with the shadow region ($\theta\approx0^\circ$) being less accurate but less important in magnitude as expected. Quantitative agreement within few percent are reached for fields 30dB down from peak value. 
	
	Next, we consider the IXV vessel case study. The working frequency is 2.26 GHz, corresponding to the standard S-band telemetry (TM) link.  
	
\begin{figure}[!ht] 
		\centerline{\includegraphics[width=0.99\columnwidth]{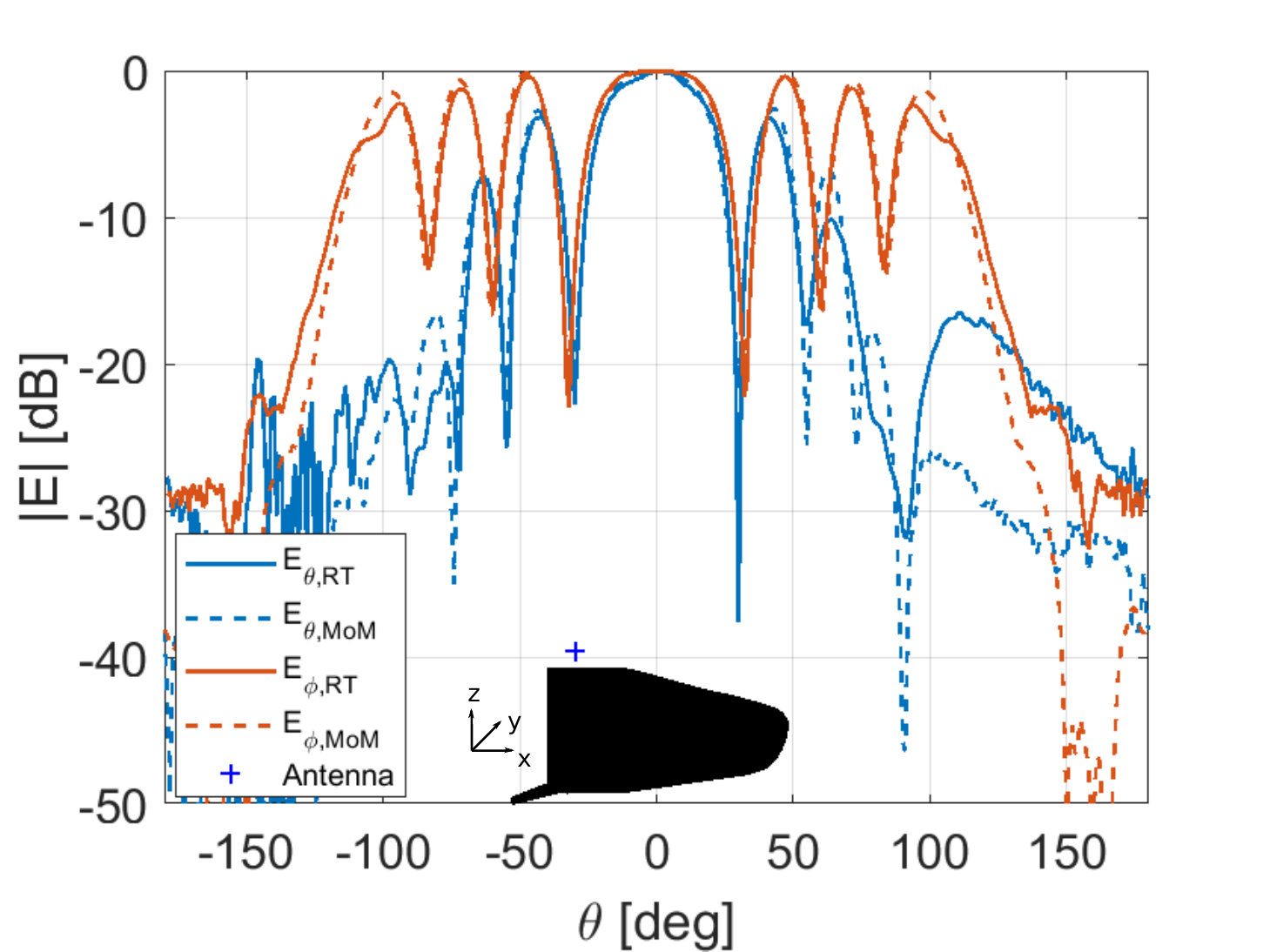}}
		\caption{Radiation of a linearly polarized source on IXV vessel; the antenna is is an x-oriented dipole at 1.77$\lambda$ distance from the surface. Cut at $\phi=0$.}
		\label{fig:benchmark:IXV_comp}
	\end{figure}

	We consider here an $x$-oriented dipole located at top rear of the vessel (see Fig.~\ref{fig:benchmark:IXV_comp}) at 1.77$\lambda$ distance from the surface; while not a typical antenna, it is considered to check the interference pattern from the interaction of the primary and reflected fields. 
	
	The cut at $\phi=0^\circ$ is plotted in Fig.~\ref{fig:benchmark:IXV_comp}. The interference pattern in both the $E_{\theta}$ and $E_{\phi}$ components are fairly well reproduced. Discrepancies are visible at large angles where fields are already more than 25dB smaller than the largest value. Note that, compared to reflection on a flat surface, the peaks are shifted laterally toward larger angles and this effect is properly captured by the simulations. Not surprisingly, it is found that a proper tessellation is crucial for a correct reflection and interference simulation for curved surfaces. The typical patch size, at least in the region nearby the source 
	should not exceed $\lambda$/5, as also typically used in MoM simulation. 
	
	\subsubsection{Scattering (RCS)}
	
	Finally we compute the differential cross-section of the ESA IXV re-entry vehicle mentioned in the Introduction; the frequency is again 2.26 GHz.
	
\begin{figure}[!ht]
		\centerline{\includegraphics[width=0.99\columnwidth]{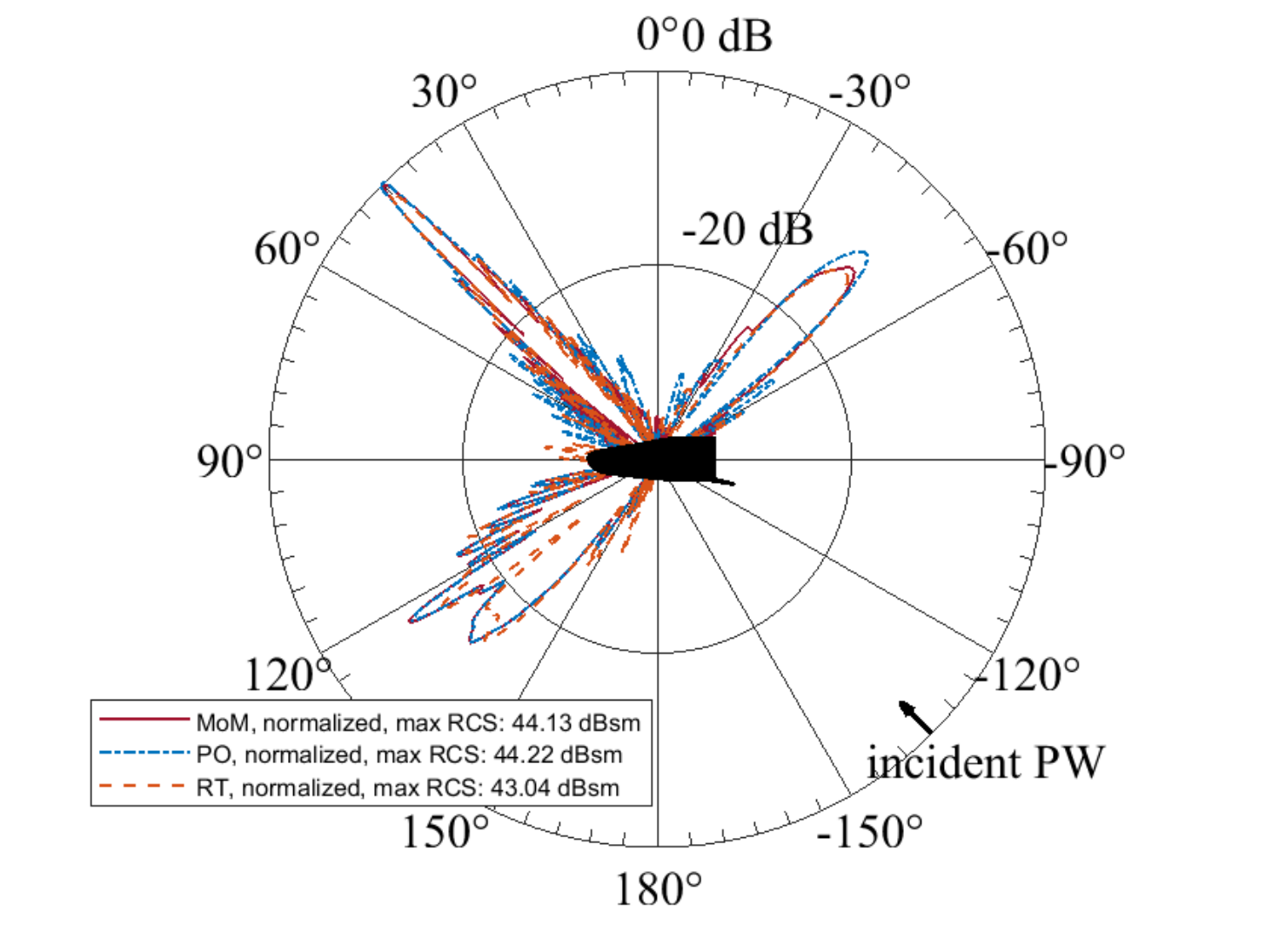}}
		\caption{IXV vessel in vacuum: bistatic RCS with incidence from $\theta=-135^{\circ}$. Comparison of MoM (full line), PO (dot-dashed line) and RT (dashed line) solutions.}
		\label{fig:benchmark:RCS_IXV_vacuum}
	\end{figure}

	We compare the Ray-tracing simulation with both MoM and PO free space calculations. 
	Incidence is in the plane of symmetry of the vehicle (mid-plane), from $\theta=45^\circ$ below the vehicle axis (see Fig.~\ref{fig:benchmark:RCS_IXV_vacuum}); the $z$axis is in the vertical direction (upward), with the $x$-axis pointing out of the drawing plane. 
	
	In Fig.~\ref{fig:benchmark:RCS_IXV_vacuum} we compare the MoM, the RT and the PO normalised scattering cross section for $\phi=0^\circ$. Three main peaks exist in the bistatic RCS profile, similar to what happens  for a finite cylinder at oblique incidence (not shown for brevity).  The peak at $\theta=45^\circ$ is the forward scattering; that at $\theta-45^\circ$ is the back scattering from (vertical) flat vehicle aft, as for the cap (disk) of a cylinder; that at $\theta=135^\circ$ is the back scattering from main body, especially the bottom of the vehicle.
	Similar results are observed for various angles of incidence of the PW, not reported for the sake of conciseness. 
	
	\section{Effect of plasma on RCS and antenna pattern: the IXV case}\label{sec:IXV_plasma}
	

	In this section we study the effect of plasma on both the scattering cross section and on the radio link for on-board antennas; we consider the plasma both including and neglecting collisional effects. 
	We consider effects on the radio links both at the S-band telemetry and control (TM, 2.26 GHz) and L-band navigation (GPS) frequencies; we study bistatic RCS at the TM frequency, as representative of the 3-5 GHz radar band.
	
	\subsection{Plane wave incidence} \label{sec:IXV_PWinc}
	
	We begin with plane-wave incidence, as it affords a simpler view of plasma-related effects.
	
	In the first place, we consider the effects of high-density layers associated with cutoff, i.e. for which the operation frequency is near or below the local plasma frequency $\omega_{pe}$ (see (\ref{eq:plasma_dispersion})). 
	
\begin{figure}[!ht] 
		\centerline{\includegraphics[width=0.99\columnwidth]{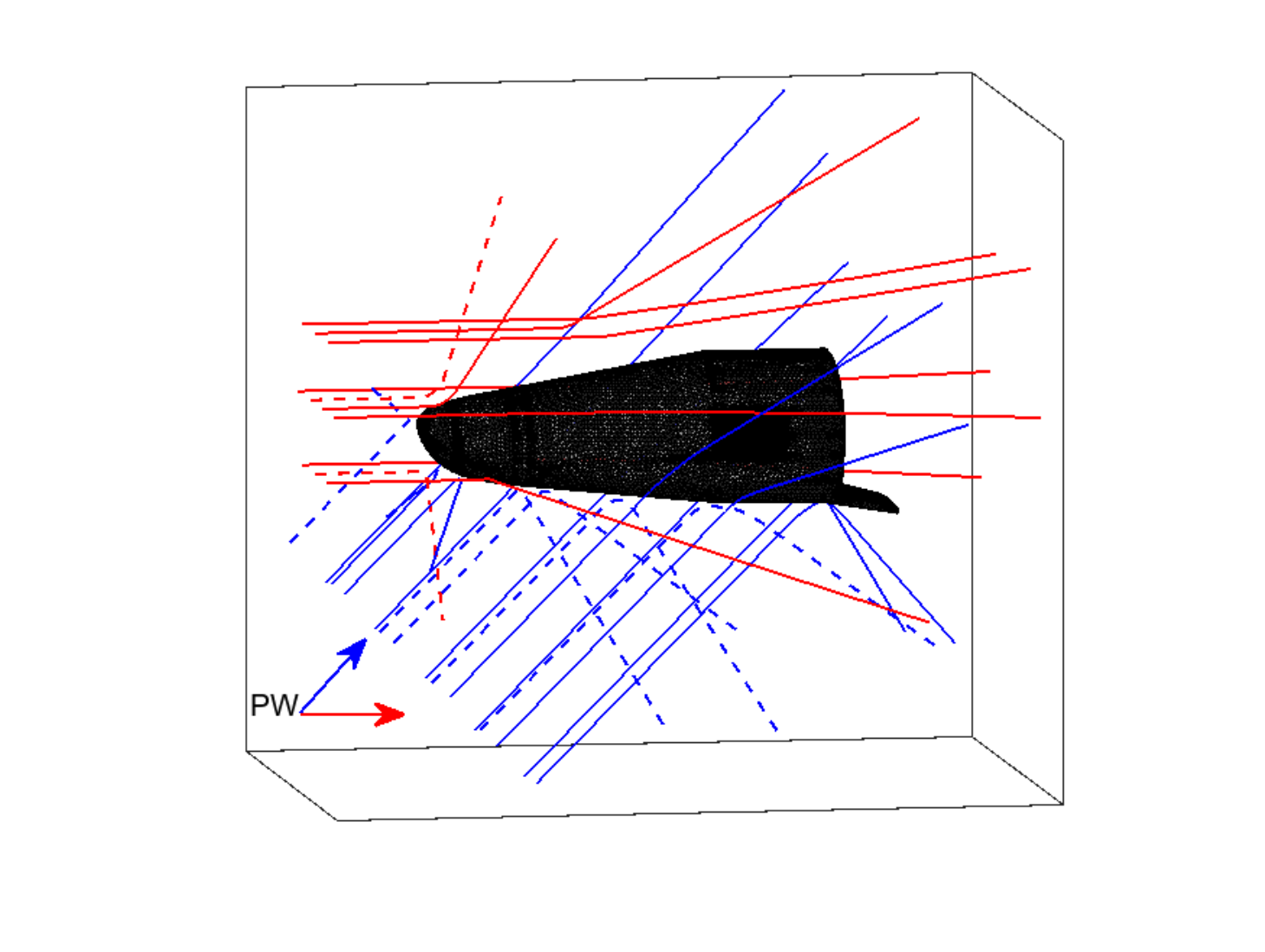}}
		\caption{IXV vehicle at Mach 12: selected ray trajectories for plane-wave illumination. The figure shows two directions of incidence. Rays in dashed lines exhibit sharp reflection-like patterns related to cut-off plasma layers underneath the vessel and in front of the nose.}
		\label{fig:plasma:Rays_mach_12}
	\end{figure}

	
	In typical re-entries, the densest plasmas are associated to thin boundary layers underneath the vehicle, and in front of the nose (as in Figs. \ref{fig:model:IXV_plasma} and \ref{fig:model:physical_model}). Fig.~\ref{fig:plasma:Rays_mach_12} depicts a small number of relevant rays (to avoid cluttering), in a flight condition where these denser layers are in cutoff at the considered frequency (TM, here). One clearly sees ``reflection-type'' effects (depicted in dashed blue and red colors for clarity), where dense layers appear to act on rays as a reflecting (e.g PEC) boundary. A closer analysis shows that rays turn around before reaching the cutoff surface, and the ``reflection'' is actually a sharp bend (see also discussion in section \ref{sec:model:validity}). In an short section of the ray path ($<\lambda$) around reflection, both density gradient and adsorption exceed the strict validity conditions and a local GO solution would result in inaccurate representation of the $\mathbf{E}$ field. We however, seek for a solution at the equivalent surface, several meters away from cut-off where a good approximation is to be expected provided we correct for the phase shift.
This explains why we did do explicitly consider such a reflection condition (which is possible, though) into the code. On the other hand, it is noted that denser layers prevent EM waves to interact with the vehicle body, and the RCS will be determined by the plasma distribution rather than the vehicle geometry. 
	
	Rays encountering lower plasma densities are also bent away from the vessel and will contribute to the scattered field in related directions.
	
	
	
\begin{figure}[!ht] 
		\centerline{\includegraphics[width=0.99\columnwidth]{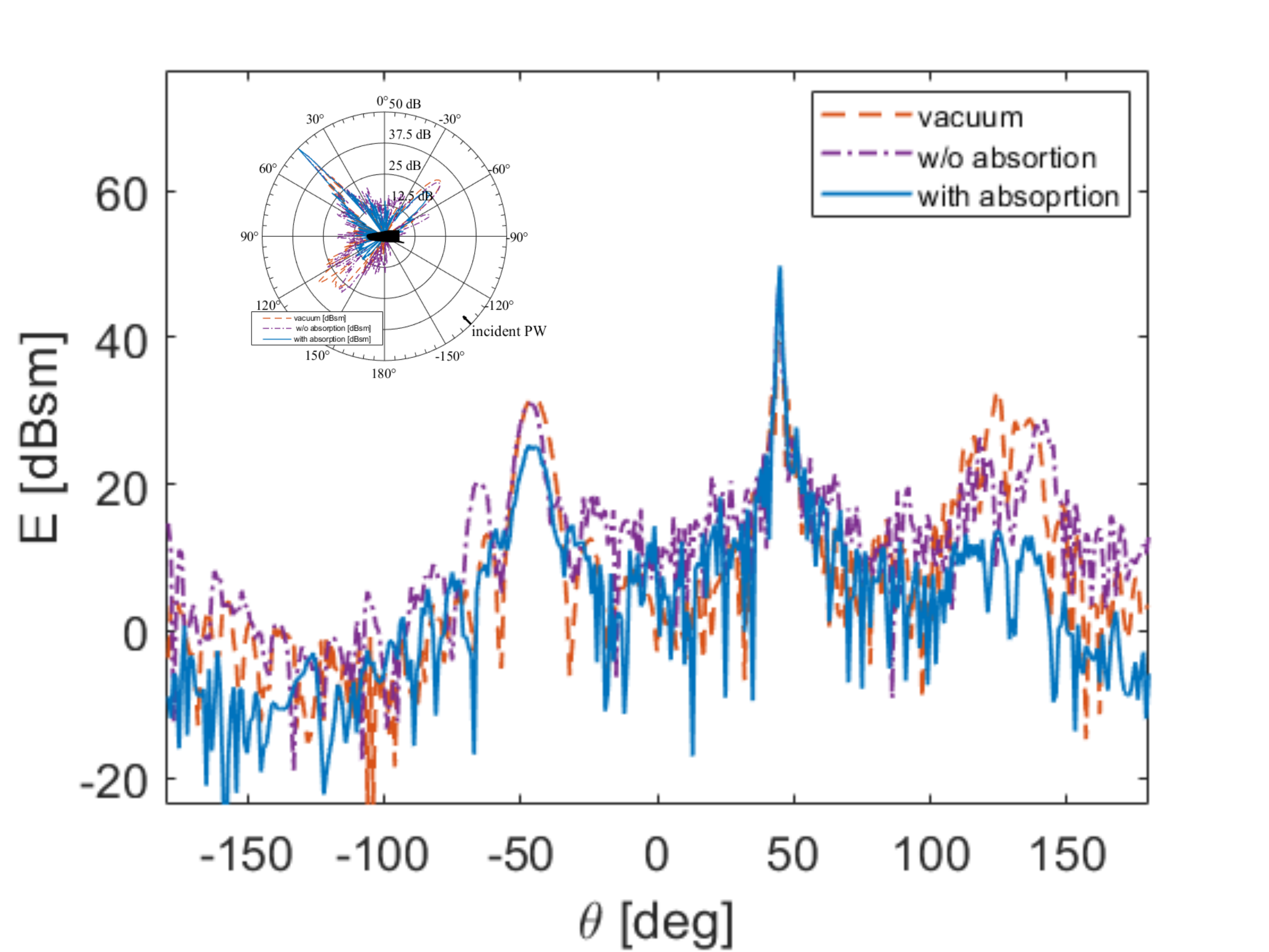}}
		\caption{Bistatic RCS of IXV vehicle at Mach 12, with incidence at $\theta=-135^{\circ}$. For ease of reference the figure shows the vacuum RCS (dashed line), along with  with the result with (with plasma power absortion) plasma (full line), the results for collisionless plasma (no absorption) are also reported (dot-dashed line). In the insert the same data is shown as polar plot.}
		\label{fig:plasma:RCS_IXV}
	\end{figure}

	The effect of the above wave phenomena on scattered field is seen in Fig. \ref{fig:plasma:RCS_IXV}; the figure compares the bistatic RCS for the vehicle alone (i.e. in vacuum) and with the plasma envelope; it also highlights the effect of plasma collisions. The ``geometric'' effect of bending or reflecting rays is apparent by comparison to the vacuum RCS: it is well-known that the forward scattering cross section arises from the masking/blockage size of the target. Since the plasma increases the apparent size of the target, this larger apparent size leads to a larger forward cross section. 
	In addition, collisional absorption acts in reducing scattered fields, as can be expected; it is to be observed that in the reported realistic conditions absorption effects are significant in RCS return.
	
	
	In a collisionless plasma (purple dashed line) the back-scattering (peak at $\theta=135^\circ$) is shifted toward higher angles and overall reduced; this is due to the ``geometric'' effect of plasma, and the peak reduction is attributable to loss of coherence. The PEC reflection is replaced by reflection-like effect from plasma around cutoff, and again the beam is more scattered than for specular coherent reflection because of inhomogeneity in the plasma volume. The other reflection peak at $\theta=-45^\circ$ is mostly unchanged in this case; this peak is due to reflection from the aft portion of the vehicle, and in this case the involved rays do not encounter a high density plasma. For higher Mach numbers, however, a high density wake plasma is formed behind the IXV vessel which impede the rays from reaching the aft flat part; in these cases (not shown)  the back-scattering peak disappears. 
	
	
	Collisional plasma absorption can be particularly effective under these conditions of both high plasma and neutral density. 
	The bistatic RCS for a Mach 12 collisional plasma (blue dash-dotted line) shows a strongly reduced back scattering.  
	The back-scattering RCS aft-related peak is reduced by several dBs but still clearly visible. For higher Mach numbers (not shown), 
	only forward radiation survives.
	
	\subsection{Radio links} \label{sec:IXV_radiolink}
	
	We now move to consider radiation from antennas mounted on the vehicle.
	
\begin{figure}[!ht] 
		\centerline{\includegraphics[width=1.0\columnwidth]{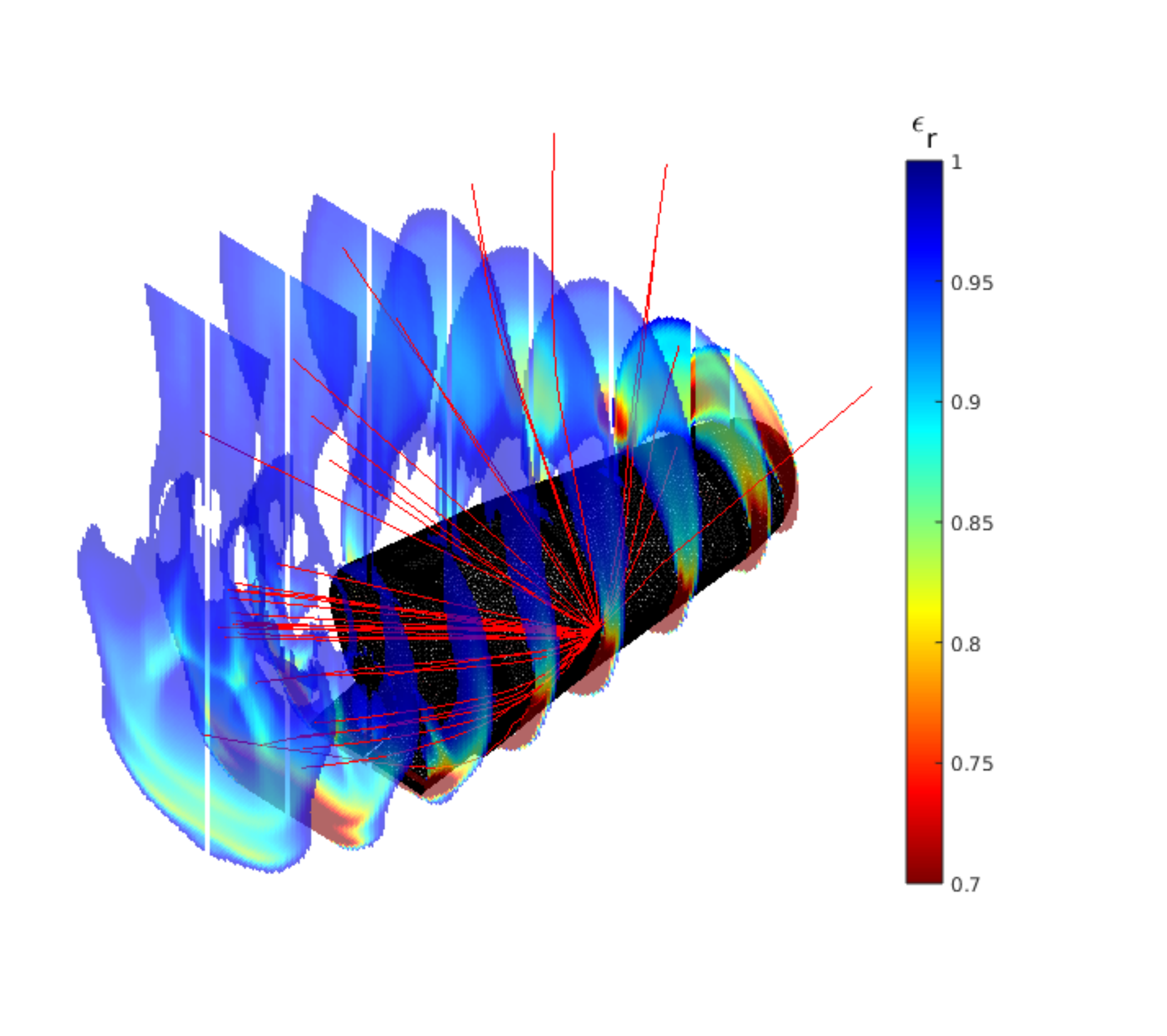}}
		\caption{TM Radiolink at Mach 15: selected ray trajectories, with plasma density overlay. Rays depart from the location of the TM antenna; plasma density is represented as relative permittivity $\epsilon_r$ at the frequency of 2.26 GHz with a color plot scale.}
		\label{fig:plasma:Rays_mach_15_antenna}
	\end{figure}

\begin{figure}[!ht] 
		\centerline{\includegraphics[width=0.99\columnwidth]{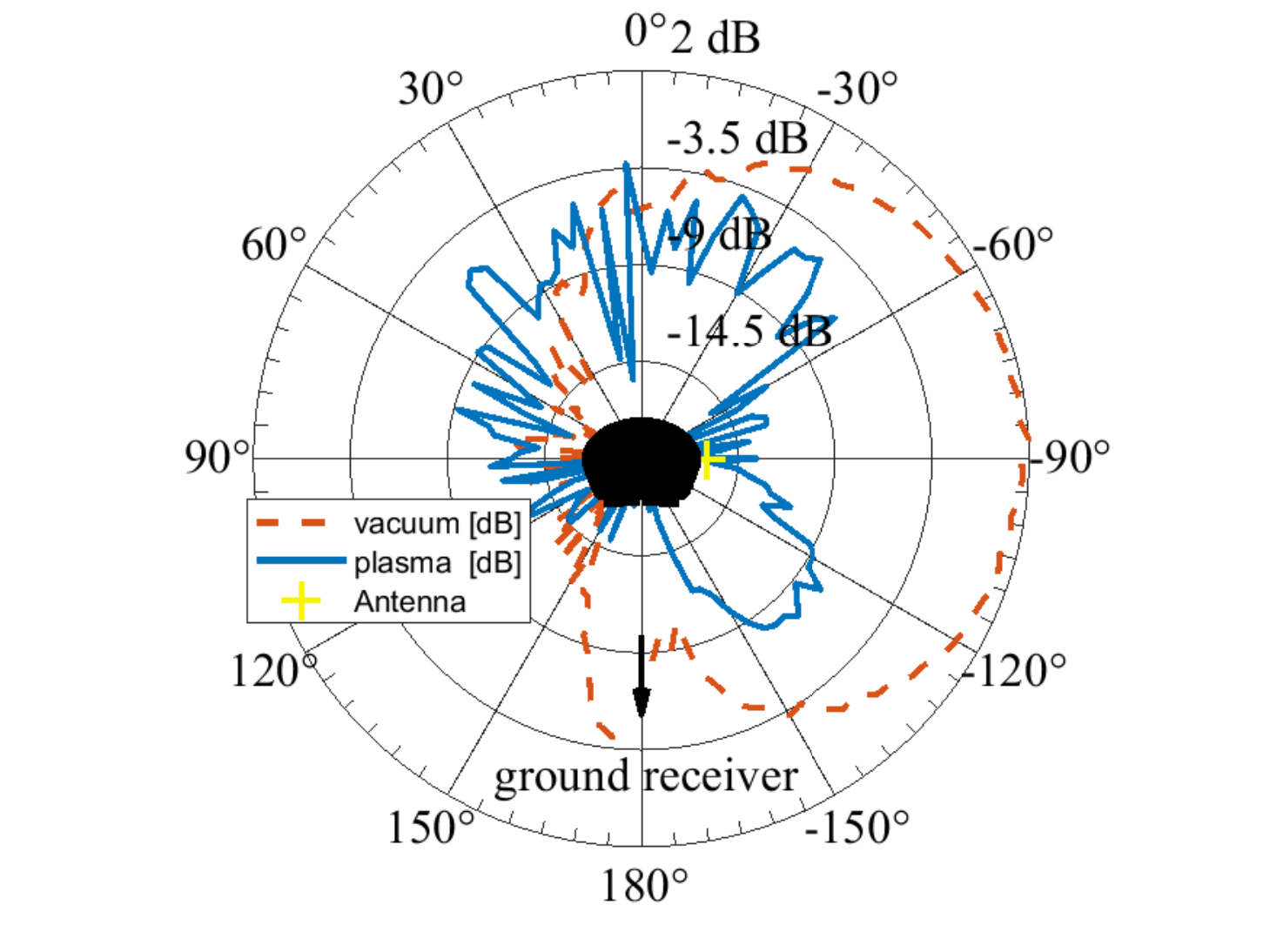}}
		\caption{TM Radiolink at Mach 15: radiation pattern of on-board antenna (co-polar component, RHCP) mounted on the right side of the IXV vehicle.  The figure shows the vacuum pattern and the pattern with plasma (with collisional absorption). The direction of the earth receiver is indicated with a black arrow.}
		\label{fig:plasma:Antenna_IXV}
	\end{figure}

	Fig. \ref{fig:plasma:Rays_mach_15_antenna} shows the relevant ray phenomena for an antenna mounted on board under the Mach 15 condition at the telemetry channel frequency (2.26 GHz). Rays depart from the location of TM antenna, on the side of the vehicle. The figure also includes some slices that illustrate the plasma's relative permittivity.
	Rays do not encounter the cut-off layer but are seen to bend toward the nearby upper region with lower plasma density (higher dielectric constant), as can be seen from the density maps. The high density and gradients are localised in a narrow region near the antenna and vessel surface. There, the normalised gradient scale length $\delta$ can reach values of 0.05 for about 10$\%$ of the rays over a path of few $\lambda$'s. Similarly, the absorption is maximal in this region, where equation \ref{eq:validity_3} becomes marginally satisfied. On most computational domain, the ray tracing assumptions are well justified and eventual error expected to be small.
	 The ensuing radiation pattern is shown in Fig. \ref{fig:plasma:Antenna_IXV}; the approximate direction of the link toward the ground station (at that point of the IXV trajectory) is indicated with an arrow. The free space radiation pattern is also reported for comparison.
	The free space pattern (red line) is strongly modified and weakened especially toward the ground direction at $\theta=180^{\circ}$ suggesting a black-out or brown-out in the communication. Interestingly, it turns out that here the link direction does not cross a cut-off layer. Hence, the strong attenuation in the radio link is not due to plasma cut-off as typically assumed; the plasma path loss is instead due to a combination of strong ray bending (seen in Fig. \ref{fig:plasma:Rays_mach_15_antenna}) and absorption.

	%

\begin{figure}[!ht] 
		\centerline{\includegraphics[width=0.99\columnwidth]{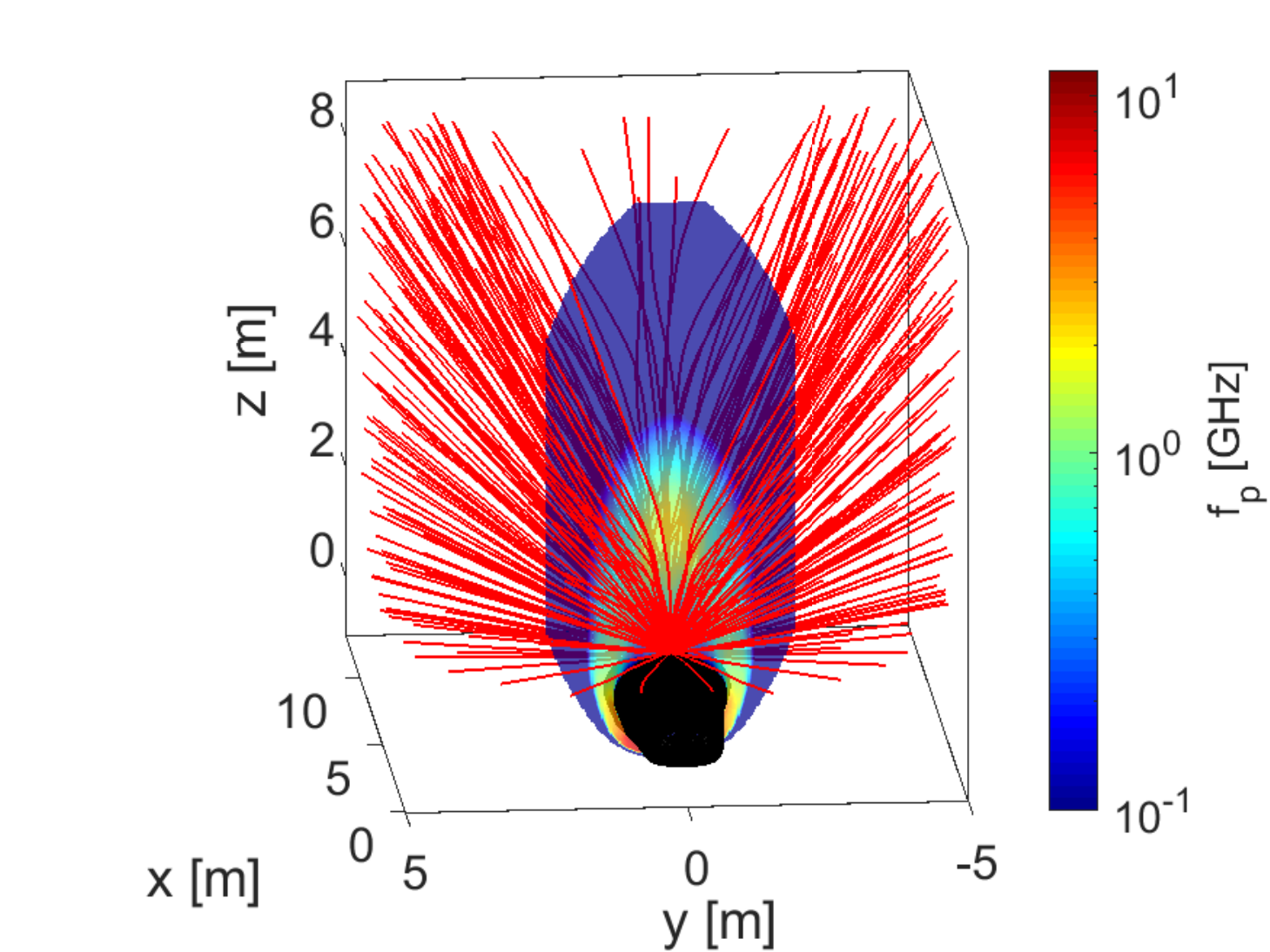}}
		\caption{GPS link at Mach 20: rays originating from the GPS antenna location.}
		\label{fig:plasma:Rays_vac_mach20_GPS}
	\end{figure}

\begin{figure}[!ht] 
		\centering
		
		
		
		\subfloat[vacuum case]{\includegraphics[width=.99\columnwidth]{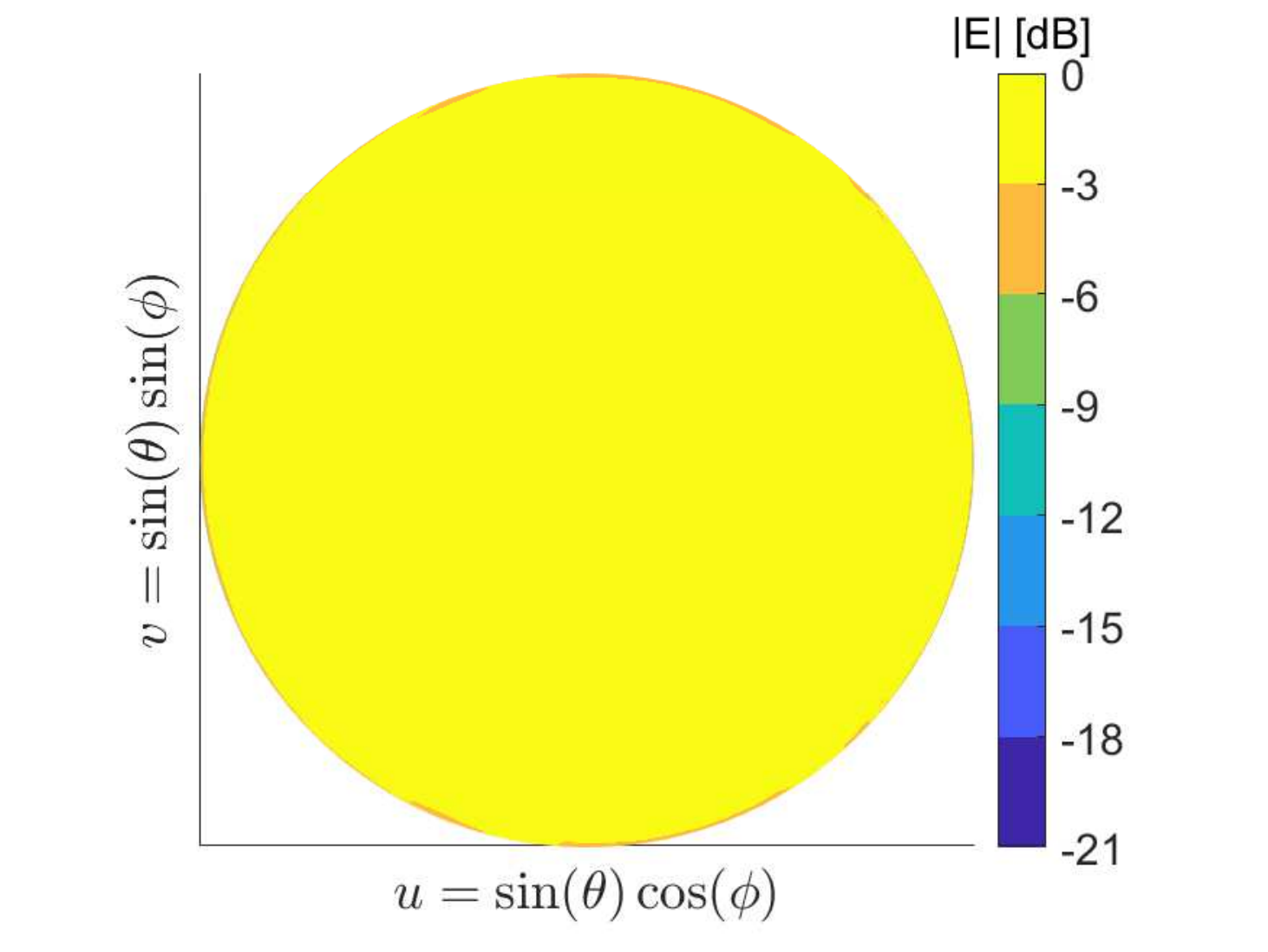}}
		
		\subfloat[plasma case]{\includegraphics[width=.99\columnwidth]{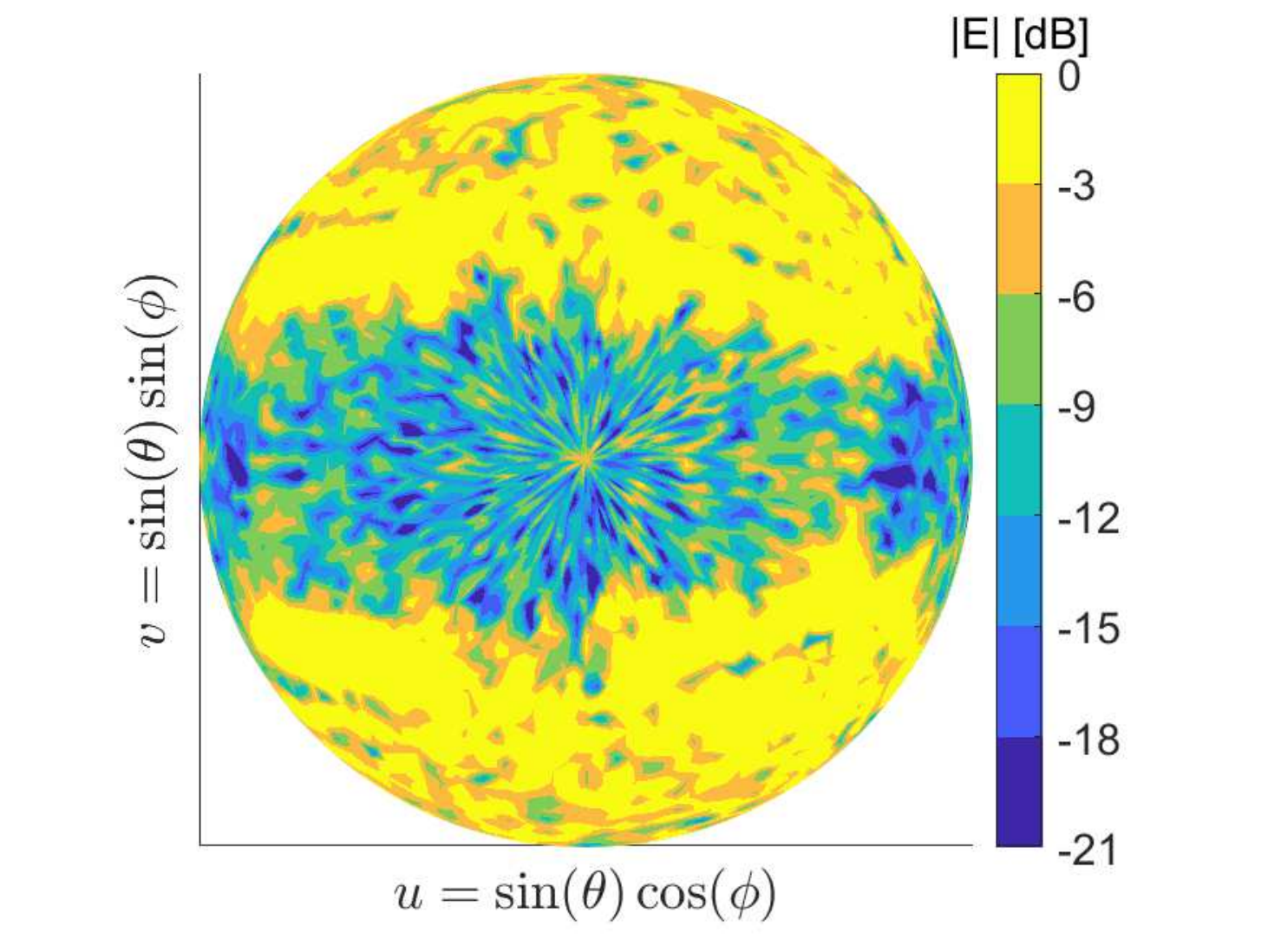}}
		
		\caption{GPS link at Mach 20: radiation pattern represented as color code on the (far field) sphere to highlight possible impact on GPS satellite links.}
		\label{fig:plasma:GPS_IXV_FarField}
	\end{figure}

	Another very interesting analysis is the impact of plasma on navigation, which employed GPS at 1.6GHz in IXV. The GPS link differs from the TM link for the (slightly) different frequency, and especially for the different antenna location, here on the top part of the vehicle. Ray analysis is shown in Fig. \ref{fig:plasma:Rays_vac_mach20_GPS}, and the resulting patterns are shown in Fig. \ref{fig:plasma:GPS_IXV_FarField}. The entire orbit of the IXV experiment is below 100km altitude, so that mostly the upper hemisphere is relevant to assess possible link obstruction; 
	results in Fig.~\ref{fig:plasma:GPS_IXV_FarField} are then displayed on the $(u, v)$-plane, with $u=\sin(\theta)\cos(\phi)$ and $v=\sin(\theta)\sin(\phi)$.
	
	While the higher density plasma layers are in the bottom part of the vehicle, the plasma is dense and inhomogeneous enough in the entire volume around as to significantly impact on extended angular regions or the antenna pattern.
	This strong modification of the pattern is seen to be primarily due to strong ray deflection upwards (Fig. \ref{fig:plasma:Rays_vac_mach20_GPS}). In turn, this deflection is due to sustained density gradients for a significant spatial region (see map), not to near-cutoff effects.
	
	
	\subsection{Numerical workload and resources}
	In the real-life cases in this section, computational domains where typically in excess of a 10m $\times$ 10m
	$\times$ 5m box, i.e. 100$\lambda$ $\times$ 100$\lambda$  
	$\times$ 50$\lambda$ at the operational frequency of 2.6GHz. Despite these very large volumes, memory needs are very modest and dominated by the input data such as the CFD plasma grid and the vehicle surface mesh. Indeed, for all simulations reported here we used a standard PC with Intel core-i7 multicore CPU and 32GB of RAM.
	
	In this proposed hybrid scheme most of the computational (CPU) time is spent by ray-tracing in the inhomogeneous plasma, whilst the radiation part only accounts for few percent of the total CPU time. For the presented cases of complex vehicle and typical sub-wavelength resolution at the Huygens surface (see section \ref{sec:model:raytube}) we trace up to 0.5M rays; this requires up to 10-12h on the indicated machine. 
	
	We note that ray tracing is very easily parallelized (and so is radiation); this, associated with the low memory requirements, makes the scaling of this approach very favorable towards multi-core or parallel machines like GPUs.
	
	
	
	\section{Conclusions}
	We have presented a method to compute wave scattering and radiation for electrically large objects immersed in a plasma, as the one originating from hypersonic flight conditions. We employ the Eikonal approximation in the large inhomogeneous plasma region, and compute radiation and scattering via Equivalence Theorem. The implementation has been benchmarked against analytical, PO and MoM references. 
	We have employed the approach to analyse bistatic RCS and radio links for a real-life re-entry vehicle, with 3D plasma profiles imported from state of the art fluid dynamic (CFD) simulations. This case study has highlighted the impact of plasma spatial variations, not only its density, on the electromagnetic response; collisional effects have likewise been found important.
	
	Future work regards extension of this method to hybrid Ray-MoM approaches or both installed antennas and for the treatment of fast gradients.
	
	\section*{Acknowledgment}
	This work was supported in part by ESA-ESTEC under contracts No. 4200022989/09/NL/JK ``Re-Entry Vehicle Communication Technology'', and 4000112664/14/NL/MH ``Mitigation of RF Blackout for Re-Entry Vehicles''.

	\bibliography{Biblio_RT} 
	\bibliographystyle{ieeetr}
		
	
\end{document}